\documentclass[letterpaper,review,authordate]{elsarticle}

\usepackage{amssymb}

\begin{document}

\begin{frontmatter}

	\title{On the Progenitors of Type Ia Supernovae\tnoteref{t1}}
	\tnotetext[t1]{To appear in \textit{Physics Reports}}

	\author{Mario Livio\fnref{1,2}}

	\fntext[1]{Department of Physics and Astronomy, University of Nevada, Las Vegas, 4505 South Maryland Parkway, Las Vegas, NV 89154, USA}
	\fntext[2]{Visiting Scholar, Department of Particle Physics and Astrophysics, Faculty of Physics, The Weizmann Institute of Science, Rehovot 76100, Israel}

\author{Paolo Mazzali\fnref{2,3}}
	\fntext[3]{Astrophysics Research Institute, Liverpool John Moores University, IC2, Liverpool Science Park, 146 Brownlow Hill, Liverpool L3 5RF, UK}

	\begin{abstract}
We review all the models proposed for the progenitor systems of Type~Ia supernovae and discuss the strengths and weaknesses of each scenario when confronted with observations. We show that all scenarios encounter at least a few serious difficulties, if taken to represent a comprehensive model for the progenitors of all Type~Ia supernovae (SNe~Ia). Consequently, we tentatively conclude that there is probably more than one channel leading SNe~Ia. While the single-degenerate scenario (in which a single white dwarf accretes mass from a normal stellar companion) has been studied in some detail, the other scenarios will need a similar level of scrutiny before any firm conclusions can be drawn.
	\end{abstract}
	
	\begin{keyword}
		\texttt{Type~Ia supernovae; white dwarfs; binary stars; thermonuclear detonation; thermonuclear deflagration}
	\end{keyword}
	
\end{frontmatter}

\section{Introduction and Motivation\label{sec1}}
Type~Ia supernovae (SNe~Ia) represent some of the most dramatic explosions in the universe, producing a luminosity of about $10^{43}$~erg~s$^{-1}$ near maximum light. They are identified by the absence of both hydrogen and helium in their spectra, and by the presence of broad (blueshifted) signatures of silicon, calcium and iron. In the early phases the lines from iron-peak and intermediate elements are typically superimposed on a thermal continuum. In late phases the spectra are characterized by iron-peak elements' forbidden lines (see e.g., Fillippenko 1997 and Branch \& Wheeler 2017 for detailed reviews of the observational characteristics). They occur in all types of galaxies, young and old.

Generally, the light curves of SNe~Ia rise to a maximum within about ten to twenty days (e.g., Zheng \& Filippenko  2017). This is followed by a decline which is more rapid in the first month (by about a factor of 15 in luminosity, or 3~magnitudes), and then a steady decline (by about 1~magnitude per 100 days).

Typically, SNe~Ia were found to exhibit a clear correlation between their peak luminosity, the rate of decline after maximum light, and the color at maximum. This so-called Phillips relationship (Phillips 1993; Phillips et~al.\ 1999; Kattner et~al.\ 2012) states that the maximum intrinsic $B$-band magnitude is given by
\[
M_\mathrm{max}(B) = -21.726 + 2.698 \Delta m_{15}(B)~,
\]
where $\Delta m_{15}(B)$ denotes the decline in the $B$-magnitude light curve from maximum light to the magnitude 15~days after $B$-maximum. Qualitatively, more luminous SNe decline more slowly and therefore have broader light curves (meaning that they radiate more energy throughout their evolution), and they are slightly bluer. The relation has been recast to include the evolution in other bandpasses, and is sometimes expressed through a stretch parameter in the time axis, relative to a standard template (e.g., Riess, Press \& Kirshner 1996; Goldhaber et~al.\ 2001). Since the light curve behavior is  independent of distance, the width-luminosity relation (or its equivalents) can be used to infer the SN's intrinsic luminosity, thereby opening the door for the use of SNe~Ia as standardizable candles. Other standardization techniques have also been developed (e.g., Burns et~al.\ 2014). 

Broadly speaking, the light curve of SNe~Ia is primarily determined by the mass of the radioactive $^{56}$Ni synthesized in the explosion, since it is the decay of $^{56}$Ni that powers the light curve. At the same time, the production of iron-group elements increases the opacity, thereby slowing down the light curve development (e.g., Arnett 1982; H\"oflich \& Khokhlov 1996; Kasen \& Woosley 2007) and affecting the peak luminosity (e.g., Pinto \& Eastman 2000a,b). There are indications that the time from explosion to peak is longer for SNe~Ia that are more luminous (e.g., Contardo, Leibundgut \& Vacca 2000).

Four simple observational facts have formed the basis for the theoretical investigations into the nature of SNe~Ia explosions:
\begin{enumerate}[1.]
\item The typical specific kinetic energy of the explosion (corresponding to ejecta moving at $\sim$$10^4$~km~s$^{-1}$) is of the order of that expected from the transformation of carbon and oxygen into $^{56}$Ni. In fact, the total energy released in the nuclear burning of a few tenths of a solar mass of carbon and oxygen into $^{56}$Ni is of the order of $\sim$$1.5\times10^{51}$~ergs, which exceeds the gravitational binding energy of a white dwarf ($\sim$$5\times10^{50}$~ergs). The remaining energy is the kinetic energy of the SN. Colgate and McKee (1969) have further shown that the observed light curve can be explained by the reprocessing of the gamma rays obtained from the radioactive decays of $^{56}$Ni to $^{56}$Co followed by $^{56}$Co decaying to $^{56}$Fe (radioactive $^{56}$Ni has a half life of 6.1 days; $^{56}$Co has a half life of 77.7 days). Hence, the luminosity and light curve of SNe~Ia are primarily determined by the amount of radioactive $^{56}$Ni synthesized in the explosion, and by the opacity of the SN ejecta, as optical photons need to escape. In addition, Mazzali et~al.\ (2007) have shown that the spectral evolution of SNe~Ia is also consistent with that expected from the presence of a few tenths of a solar mass of $^{56}$Ni and that the observations suggest that many SNe~Ia may be associated with an exploding object of a similar mass.

\item The event is violently explosive---pointing to a runaway process, that could plausibly be associated with degenerate conditions in which the pressure is almost independent of temperature and therefore cannot act as a regulating ``valve'' as the temperature rapidly rises (e.g., Arnett 1982; Nomoto, Thielemann \& Yokoi 1984; Woosley \& Weaver 1986).

\item The lack of hydrogen and helium in the spectrum indicates that the exploding star is not a normal (main sequence) star. In addition, observations of SN~2011fe implied a size of $R_*\lesssim0.1~R_\odot$ for the exploding star (Bloom et~al.\ 2012; Piro \& Nakar 2012), suggesting either a degenerate star or at least an unusual star (such as a carbon star).

\item SNe~Ia explosions are observed in both young and old stellar populations, sometimes with long delays after the cessation of star formation (e.g., Mannucci, Della Valle \& Panagia 2006).
\end{enumerate}

\begin{figure}
\centering
\includegraphics[width=.9\linewidth]{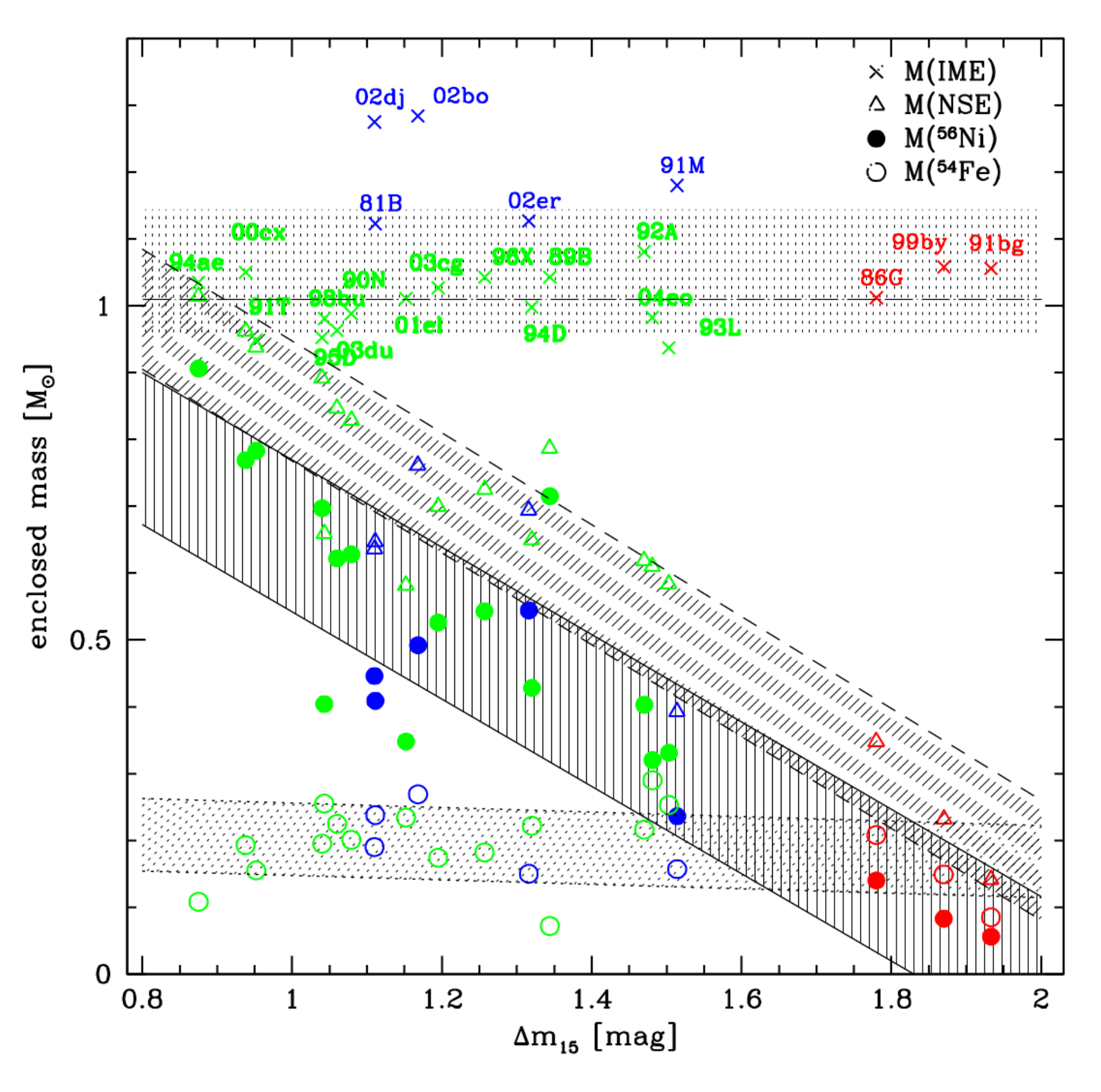}
\caption{Distribution of the principal isotopic groups in SNe~Ia. The enclosed mass (obtained from spectral modeling) of different burning products is shown versus decline rate parameter $\Delta m_{15}(B)$ (a proxy for SN luminosity). Individual SNe are colored according to their velocity evolution (Benetti et~al.\ 2005): high velocity gradient (HVG), blue; low velocity gradient (LVG), green; and faint, red. Open circles indicate the mass of stable $^{54}$Fe + $^{58}$Ni for each SN, solid circles indicate that of $^{56}$Ni, and open triangles indicate the sum of these(total NSE mass). Crosses show the sum of NSE and IME mass, indicating the total mass burned. The IME mass is the difference between crosses and triangles. $^{54}$Fe and $^{58}$Ni are found in roughly constant amounts in the deepest parts of all SNe, irrespective of luminosity: $M$(stable NSE)$=0.24-0.03\Delta m_{15}(B)~M_{\odot}$, with rms dispersion 0.05~$M_{\odot}$ (lower horizontal shaded area). The $^{56}$Ni mass determines the SN luminosity. It correlates with $\Delta m_{15}(B)$: $M(^{56}Ni)=M(NSE)=1.54-0.69\Delta m_{15}(B)~M_{\odot}$, rms dispersion 0.09~$M_{\odot}$ (upper diagonal shaded area). IMEs lie mostly outside the iron-group zone. The outer Si velocity is similar for all SNe except HVG SNe. The mass enclosed by IMEs represents the total burned mass. When all SNe are included, the average value is $\sim$$1.03\pm0.09~M_{\odot}$ (upper shaded area). Both values are almost independent of $\Delta m_{15}(B)$.}
\end{figure}
	
Additional information can be deduced from Figure~1 (adapted from Mazzali et~al.\ 2007, with a few new SNe and new studies added), which shows the composition of different SNe~Ia ordered by decline rate. Empty circles represent the mass of stable nuclear-statistical-equilibrium (NSE) elements derived from nebular modeling. Filled circles represent the $^{56}$Ni mass derived from nebular modeling as well as light curve fitting. Triangles show the sum of these two components, which is the quantity that most directly affects the opacity and therefore the light curve (LC) shape. Finally, crosses indicate the extent of Silicon as representative of incomplete burning to intermediate-mass elements (IME), which produces no light but contributes almost as much as burning to Fe-group to creating kinetic energy. The plot shows that for most SNe the amount of total mass burned is similar. This also implies that the kinetic energy is rather similar in all events, because IME replace $^{56}$Ni. Such a configuration suggests delayed detonation explosions in a Chandrasekhar-mass WD (see discussion in Section~\ref{sec2.2}).

Taken together, the four points above are the main reasons that have led to the generally accepted model for the event that is responsible for SNe~Ia explosions: \textit{the thermonuclear disruption of a white dwarf star}. This consensus model received a resounding confirmation when observations of SN~2014J detected $^{56}$Co lines at energies of 847 and 1,238 KeV, with the line fluxes suggesting that about $0.6\pm0.1~M_\odot$ of radioactive $^{56}$Ni had been synthesized in the explosion (Churazov et~al.\ 2014, 2015; Diehl et~al.\ 2014).

The detailed physics of the ignition of nuclear reactions and the propagation of the ``flame'' of nuclear burning through the white dwarf (WD) material still involve a number of uncertainties. In particular, it is not entirely clear whether the explosion proceeds through a pure detonation (in which the burning front propagates faster than the speed of sound; e.g., Kushnir et~al. 2013), or whether (and how) it involves a transition from deflagration (flame propagating at subsonic speeds) to detonation (see e.g., Nomoto 1982; Khokhlov 1991; Woosley et~al.\ 2009; Hillebrandt et~al.\ 2013 for detailed discussions). The ``classical'' arguments have been that a direct detonation may produce more iron-peak elements than would be in agreement with galactic chemical evolution but would not produce intermediate mass elements, while a pure deflagration would not produce a sufficient amount of $^{56}$Ni to power the light curve (e.g., Woosley, Taam \& Weaver 1986; H\"oflich, Wheeler \& Thielemann 1998). These topics, however, while very important and interesting in their own right, are somewhat beyond the main scope of the present review, which will concentrate primarily on the question of the \textit{progenitors} of SNe~Ia. 

There is another observational characteristic of SNe~Ia that should be mentioned because it could provide important clues about the progenitor systems---the Delay Time Distribution (DTD). The idea is to determine the distribution of times between the formation of the progenitor systems and the SN explosions, by observing the SN rate as a function of the age distribution of the parent population. Since, as we shall see, different progenitor scenarios predict (at least in their simplest formulations) different DTDs, an observational determination of the DTD can, in principle at least, support or rule out progenitor scenarios (Maoz \& Mannucci 2012).

Attempts to determine the DTD have been based on: (a)~the SN~Ia rates per unit mass in clusters of galaxies and in field elliptical galaxies, (b)~the iron abundance in galaxy clusters (since SNe~Ia are the main producers of iron), (c)~a comparison of the volumetric SN rate from field surveys, as a function of redshift, to the cosmic  star-formation history, (d)~a comparison of the \textit{galactic} volumetric SN rate (for individual galaxies) as a function of redshift to the galactic star-formation history.

\textit{Most of these observations have been consistent with a DTD that is proportional to} $t^{-1}$ (see Maoz, Mannucci \& Nelemans 2014 for an extensive discussion and references therein). While a recent study based on a comparison of observed and predicted color distributions of SNe~Ia hosts from two supernova surveys found for old progenitor systems a somewhat steeper slope for the DTD, of $-1.50^{+0.19}_{-0.15}$ (Heringer et~al.\ 2017), this study made more simplifying assumptions (e.g., detection efficiencies were not considered) than in the work of Maoz, Mannucci \& Brandt (2012).

Finally, we should mention that the supernova remnants (SNRs) of SNe~Ia tend to be quite spherically symmetric, with a few showing some hints of axial symmetry.

\subsection{Why is identifying the progenitor systems so important?} 
There are at least six main reasons that make the identification of the progenitors of SNe~Ia significant:
\begin{enumerate}[(i)]
	\item Type Ia supernovae represent a fascinating phenomenon that is the outcome of stellar evolution, the evolution of binary (and perhaps even triple) systems, and thermonuclear physics. Determining the precise nature of the progenitors of SNe~Ia can elucidate various evolutionary phases and may provide valuable insights into far from fully understood processes such as stellar mass loss, common envelope evolution, and accretion onto white dwarfs. 
	
	\item Observations of SNe~Ia have led to the dramatic discovery that the cosmic expansion is accelerating (Riess et~al.\ 1998; Perlmutter et~al.\ 1999). 	SNe~Ia continue to be a key ingredient in the attempts to determine the equation of state (the ratio of pressure to density) of the ``dark energy'' that appears to propel the acceleration (e.g., Rest et~al.\ 2014; Sullivan et~al.\ 2011). 	An accurate determination of the systematic uncertainties involved in SNe~Ia observations requires an understanding of the redshift evolution of the supernova population. This, in turn, calls (among other things) for a reliable identification of the progenitor systems.
	
	\item A recent determination of the local value of the Hubble constant, H$_0=73.24\pm1.74$~km~s$^{-1}$ Mpc$^{-1}$ (Riess et~al.\ 2016), is 3.4$\sigma$ higher than the value predicted by the Planck Collaboration  (2016) from a standard ($\Lambda$\textit{CDM}) cosmological model (with 3~neutrino flavors), based on Planck Cosmic Microwave Background (CMB) data. This tension could point to new physics (e.g., one additional neutrino species, such as a sterile neutrino, or to a time-dependent dark energy equation of state), but it could also be the result of unexplained systematic errors in the CMB measurements or in the determination of the (local) value of H$_0$ (or both). While not directly related to the nature of the progenitors, again taming supernova systematics may require a better understanding of the evolution of SNe~Ia luminosity with cosmic time (e.g., its potential dependence on the cosmic metallicity).
	
	\item SNe~Ia provide input into the interstellar medium of galaxies in the form of kinetic and radiative energy, in addition to injecting a variety of nucleosynthetic products. The latter include iron (of which SNe~Ia are the main producer), silicon, calcium, sulfur, carbon and oxygen. This means that the chemical evolution of galaxies, as well as galaxy evolution in general, are affected by the cosmic history of SNe~Ia (see e.g., Mitra, Dav\'e \& Finlator 2015). SNe~Ia probably also play an important role in accelerating cosmic rays, and they may contribute to the diffuse extragalactic gamma-ray background.
	
	\item One of the suggested models for SNe~Ia involves the merger of two white dwarfs (see Section~\ref{sec3}). Such coalescence events are potential targets for the proposed ``evolved Laser Interferometer Space Antenna/New Gravitational Wave Observatory (eLISA/NGO; e.g., Amaro-Seoane et~al.\ 2013; Kilic et~al.\ 2014). 
	The recent three detections of merging black holes with Advanced LIGO (e.g., Abbott et~al.\ 2016, 2017a) and the detection of merging neutron stars by LIGO and Virgo (Abbott et~al.\ 2017b) makes the prospects of realizing such a detection (of merging WDs) in the not-too-distant future more realistic.
	
	\item Finally, we note that the fact that after decades of intensive research, we have still not unambiguously identified the progenitors of some of the most dramatic cosmic explosions has become somewhat of an embarrassment (see e.g., reviews by Livio 2000; Hillebrandt et~al.\ 2013; Maoz, Mannucci \& Nelemans 2014).	
\end{enumerate}

While we cannot promise to altogether remedy the fact that we don't have a single agreed-upon model for the progenitors of SNe~Ia in the present article, we shall at least critically review all the progenitor models that have been suggested, and we shall confront those models with extensive, up-to-date observational data. 

The literature on SNe~Ia is vast. We have not attempted to give all the relevant references. Rather, the references mentioned in this article should merely be regarded as representative. We sincerely apologize to the many authors whose important works have not been specifically cited. For a recent, excellent review from an observational perspective see Maoz, Mannucci and Nelemans (2014), and see also the more theoretically flavored review of Wang \& Han (2012). An even more recent review that includes an extensive bibliography is Maeda \& Terada (2016).

This article is organized as follows: In Section~\ref{sec2} we briefly discuss the expected properties of the exploding WD. The various proposed scenarios for the progenitor systems of the majority of SNe~Ia are presented in Section~\ref{sec3}, together with a review of what we regard as the strengths and weaknesses of each scenario, in view of all the available observational data.  In Section~\ref{sec4} we briefly discuss diversity within the main class of SNe~Ia and some transitional subclasses, as well as minority subclasses of SNe~Ia. The minority subclasses include primarily: (1)~Overluminous SNe~Ia in which the inferred $^{56}$Ni mass suggests that the exploding WD was more massive than the Chandrasekhar mass (e.g., Howell et~al.\ 2006). (2)~SNe which have been dubbed Type~Iax, which exhibit lower line velocities than normal SNe~Ia and are fainter (e.g., Foley et~al.\ 2013). This group likely contains different types of events. (3)~SNe~Ia that show clear signs of the explosion ejecta interacting with a relatively dense circumstellar medium (CSM), sometimes referred to as SNe~Ia-CSM. A discussion and tentative conclusions follow.

\section{Properties of the Exploding White Dwarf\label{sec2}}

Since we know that SNe~Ia are produced by the thermonuclear disruption of WDs, we can now advance one step further, and explore what the implied properties of the exploding object are. Specifically, we are interested in its composition and its mass.

\subsection{Composition\label{sec2.1}}

White dwarfs can be composed of helium (He WDs), of carbon and oxygen (CO WDs), or of oxygen and neon (ONe WDs). At the low-mass end are the helium WDs. The masses of He WDs are generally smaller than about 0.45~$M_\odot$. If such WDs accrete matter, they could ignite He at their centers when the WDs reach a mass of about 0.7~$M_\odot$, leading to an explosion. However, the resulting ejecta would consist entirely of He, $^{56}$Ni, and decay products (e.g., Woosley, Taam \& Weaver 1986; Nomoto \& Sugimoto 1977). Such a composition is inconsistent with observations of SNe~Ia. Prior to maximum light, the spectra of SNe~Ia are characterized by high-velocity intermediate mass elements (Mg to Ca), and in the late, nebular phase, the spectra are dominated by forbidden iron lines (e.g., Kirshner et~al.\ 1993; Wheeler et~al.\ 1995; Ruiz-Lapuente et~al.\ 1995; G\'omez, Lopez \& Sanchez 1996; Filippenko 1997). Furthermore, as we have noted in the introduction, SNe~Ia are characterized by the absence of He in their spectra. Consequently, we have to conclude that the majority of SNe~Ia are certainly \textit{not} produced by exploding helium WD stars.

At the high-mass end lie the ONe WDs. Those are believed to form in binary systems, from main-sequence stars with initial masses of around 10~$M_\odot$ [mass transfer in a binary system allows for an upper limit somewhat higher than 8~$M_\odot$---the highest mass that a single star can have and still end up as a WD, e.g., Iben \& Tutukov (1985); Canal, Isern \& Labay (1990); Dominguez, Tornamb\'e \& Isern (1993); Ritossa, Garcia-Berro \& Iben (1996)]. Existing simulations (most of which are admittedly relatively old) suggest, however, that upon accreting mass ONe WDs tend to produce an accretion-induced collapse (AIC), leading to the formation of a neutron star, rather than to an explosion (e.g., Nomoto \& Kondo 1991; Gutierrez et~al.\ 1996; Saio \& Nomoto 1985, 2004; Sato et~al.\ 2015). \textit{If} this conclusion holds true, then SNe~Ia are also \textit{not} the result of exploding ONe WDs. We should also note that in addition to the last argument (about ONe WDs leading to AICs), it is very unlikely that ONe WDs are numerous enough to produce the observed SNe~Ia rate (e.g., Livio \& Truran 1992).

The discussion above leads us to the deduction that SNe~Ia most probably represent exploding CO WDs. This inference is supported by the fact that simulations suggest (e.g., Nomoto \& Kondo 1991; Nomoto et~al.\ 2007; Hillman et~al.\ 2016; and references therein) that for a relatively wide range of initial WD masses and accretion rates, CO WDs can (in principle at least) explode if they reach close to the Chandrasekhar limit (the maximum mass WDs can have, when supported by electron degeneracy; about 1.4~$M_\odot$). As we shall discuss in Section~\ref{sec3}, in the case of two colliding WDs, explosions can be obtained even if the individual masses are smaller than the Chandrasekhar mass. An initial explosion near the surface of a sub-Chandrasekhar white dwarf, which triggers a subsequent explosion of the bulk of the material may also produce an SN~Ia (e.g., Nomoto \& Sugimoto 1977; H\"oflich \& Khokhlov 1996).

Observationally, a CO composition for the exploding WD is supported by the layered composition of the ejecta (intermediate-mass elements in the outer layers and iron-peak elements in the center; e.g., Mazzali et~al.\ 2007), as well as by the detection of $^{56}$Co lines in SN~2014J, mentioned in Section~\ref{sec1} (Churazov et~al.\ 2014).

Perhaps the strongest evidence for the exploding WD being (at least in some SNe~Ia) of CO composition comes from detailed observations of SN~2011fe (PTF11kly) in the galaxy M101 (Nugent et~al.\ 2011). The spectra of this relatively close-by (at a distance of 6.4~Mpc) SN~Ia showed strong features from unburnt material consisting of carbon and high-velocity oxygen. Furthermore, Blondin et~al.\ (2012) detected C\,{\sc{ii}} $\lambda$6580 (and hence the presence of unburnt carbon) in 23 early-time SN~Ia spectra. We can therefore advance one step further in our attempt to close in on the nature of SNe~Ia and their progenitors, and say that they are \textit{most likely produced by exploding CO white dwarfs}.

\subsection{Mass\label{sec2.2}}

Given the following two facts, first, that the luminosities of SNe~Ia are not all precisely the same, and second, that the luminosity is primarily determined by the amount of $^{56}$Ni that is produced, a natural question arises: Can the indicated range in $^{56}$Ni masses (0.1--1~$M_{\odot}$) be taken to mean that the exploding star's mass also varies from one SN to another? In particular, it may be tempting to assume that the mass of the $^{56}$Ni that is synthesized is (approximately) proportional to the WD mass. 

In order to attempt to answer this question we have to first examine proposed subclassifications of SNe~Ia, so as to determine to what extent SNe~Ia (or at least some part thereof) form a homogeneous class.

One popular spectroscopic subclassification, based on the equivalent widths of absorption features near 5750~\AA\ and 6100~\AA, was suggested in a series of papers by Branch et~al.\ (2005, 2006, 2007, 2008, 2009, and references therein). The maximum light spectra were divided into four groups: \textit{core normal} (CN), \textit{broad line} (BL), \textit{cool line} (CL), and \textit{shallow silicon} (SS). SN~2006x, for example, which was classified as extreme BL, has a very broad 6100~\AA\ absorption, but only a weak 5750~\AA\ absorption. The 6100~\AA\ absorption of SN~1991bg, classified as an extreme CL, is comparable to that of SN~1998bu, classified as a CN, but the 5750~\AA\ absorption is strong. SN~1991T, classified as extreme SS, has only shallow 6100~\AA\ absorption and hardly any 5750~\AA\ absorption. We should note though that Branch et~al.\ (2009) suggested that for the most part, the spectra appeared to have a continuous distribution of properties, rather than breaking up into discrete subgroups (with a few possible exceptions).

A plot of the equivalent width of 5750~\AA\ against the equivalent width of 6100~\AA\ reveals a dense concentration of the ``core normals,'' with the members of the other groups being more broadly dispersed (although similar in numbers; Blondin et~al.\ 2012). Core normals also show a quite remarkable homogeneity in their spectra (in the range from 4000~\AA\ to 7000~\AA) at post-maximum epochs. It is interesting, therefore, to explore whether this spectral homogeneity also translates to the peak luminosities of SNe~Ia. Figures~13 in both Branch et~al.\ (2009) and Blondin et~al.\ (2012) show the distribution of the four groups on the Phillips-relation (absolute magnitude vs.\ rate-of-decline) plane. The figures demonstrate a few interesting characteristics: (1)~On average, the SSs (of which the very luminous SN~1991T and SN~2000cx are members) are brighter than the CNs and decline more slowly (low values of $\Delta m_{15}$). (2)~The CLs (of which the faint SN~1991bg is a member) are rapidly declining and faint. (3)~On average, BLs (of which SN~2006X is a member) decline faster than CNs.

The general impression one gets from these figures is that SNe~Ia may represent a continuous distribution, but one cannot rule out the possibility that one (or more) subgroup is physically discrete. In terms of the statistics, about 25\% of all SNe~Ia can be classified as core normals (see also Ashall et~al.\ 2016). In the following discussion, (both regarding the WD mass and in attempting to identify the progenitors) we shall concentrate more on core-normal SNe~Ia, even though we'll discuss the other subclasses as well (Section~\ref{sec4}). More precisely, we shall first focus our attention on those SNe~Ia that show a relatively high degree of homogeneity in their characteristics, in order to examine how far we can get in our attempt to identify the progenitor systems for that class.

In addition to the spectroscopic homogeneity, there are a few other observational facts and theoretical considerations that seem to suggest that the masses of the exploding WDs in at least a significant fraction of the SNe~Ia are nearly the same.

First, WDs of different masses have rather different central densities. The thermonuclear burning rate is a very sensitive function of the density, with the lower densities (corresponding to smaller WD masses) being characterized by much slower rates, and concomitantly producing disproportionately smaller amounts of $^{56}$Ni. This appears to be in contrast with the observations, which seem to imply a smooth distribution of $^{56}$Ni masses, strongly peaked around 0.5--0.7~$M_\odot$ (e.g., Mazzali et~al.\ 2007). As we shall discuss in Section~\ref{sec3.4}, the central density argument may have to be revisited if SNe~Ia are caused by direct collisions (induced by a third body) of two WDs.

Second, all ``normal'' SNe~Ia show both Fe and Si in their spectra. Iron is produced when burning is complete  to nuclear statistical equilibrium (at higher densities), either to $^{56}$Ni or to stable Fe directly. Silicon is the most abundant intermediate-mass element (IME; the group that includes S, Ca, and Mg). These IMEs are produced most readily when burning is incomplete (meaning that the high densities necessary for further burning are not reached), either because of expansion or because of intrinsically low densities in the outer layers of the progenitor, or both. The point is that observations of the velocity distributions of these elements in SNe~Ia seem to reveal that Fe and Si are well separated in velocity space, with Fe being associated with lower velocities. This inferred stratification becomes even more convincing in detailed spectral models (e.g., Stehle et~al.\ 2005; Tanaka et~al.\ 2011). When the mass density of different elements is recovered from spectrum synthesis work, an important result emerges: for ``normal'' (or even slightly under- or over-luminous SNe~Ia), the mass in intermediate-mass elements almost precisely compensates for the differences in $^{56}$Ni mass (e.g., Mazzali et~al.\ 2007; Sasdelli et~al.\ 2014). This suggests at least that the outermost layers that undergo burning do so under similar conditions in all normal SNe~Ia, regardless of the production of $^{56}$Ni. Basically, models seem to imply that the mass that is nuclearly processed to elements heaver than oxygen is roughly constant in many normal SNe~Ia, with a value of about 1.0--1.2~$M_\odot$. Since the outermost layers are harder to explore (because very early data are needed), and because some oxygen does survive unburnt in most SNe~Ia, the above conclusion suggests that the masses of the progenitor WDs in many SNe~Ia may not only be roughly constant, but that they are also close to the Chandrasekhar mass. A similar conclusion is obtained from the fact that $^{56}$Ni appears to be distributed spherically, with the stable iron-peak elements being distributed with an offset of about 2000~km~s$^{-1}$ (Maeda et~al.\ 2010). 

It is important to note that while the work of Mazzali et~al.\ (2007) used the W7 explosion model which was based on a Chandrasekhar mass (Nomoto et~al.\ 1984), the results did not strongly depend on the use of this particular model. The only feature which was influenced by the use of W7 was deriving an Si boundary mass from the velocity. This dependence, however, is not strong. As we shall discuss later, however, other work suggests that there may be a fraction of normal SNe~Ia which contained WDs with masses smaller than the Chandrasekhar mass (Scalzo et~al.\ 2014; Scalzo, Ruiter \& Sim 2014).

When viewed as a sequence, much of Branch's classification simply looks at properties of SNe with different degrees of burning. For example, SS are on average more luminous than CN. As they produced more $^{56}$Ni, they produced less IME, which means that the Si line will be shallower, both because of the smaller production and the higher temperature. Conversely, CLs are SNe that produce less $^{56}$Ni, have lower luminosity, and therefore lower temperature. Nugent et~al.\ (1995) showed nicely how the temperature affects the Si\,{\sc{ii}} lines and Hachinger et~al.\ (2008) explained this somewhat unexpected behavior. BL have not been studied in detail. They tend to coincide with normal SNe, which is also the region where high-velocity features (HVF) are stronger. It is quite possible that BL-type SNe~Ia just represent strong HVF (affecting the Si line, as discussed above).

Only the SN~1991bg subclass (which mostly overlaps with Branch's CL class), may be an effectively separate class. Figure~2 shows the luminosity distribution of observed SNe~Ia in $M(B)$ and $\Delta M_{15}(B)$. These are not luminosity functions because they are not complete samples, but they represent the distribution of observed SNe only, within $z=0.06$ (see Ashall et~al.\ 2016 for the origin of the data). The main population includes Branch's CN, BL, and SS, where SS are at the luminous end. At the low-luminosity-fast-declining end, CL correspond to SNe of the 1991bg type. This group shows an upturn in number which may indicate a different population. They are also separated in properties which also suggest a separate population (see also Dhawan et~al.\ 2017).  This group includes the so-called ``transitional'' SNe as well, which bridge the gap from normal to 1991bg (see, e.g., Gall et~al.\ 2017). As we mention later, this group may include SNe that come from one or more different channels.

\begin{figure}
\centering
\includegraphics[width=\linewidth]{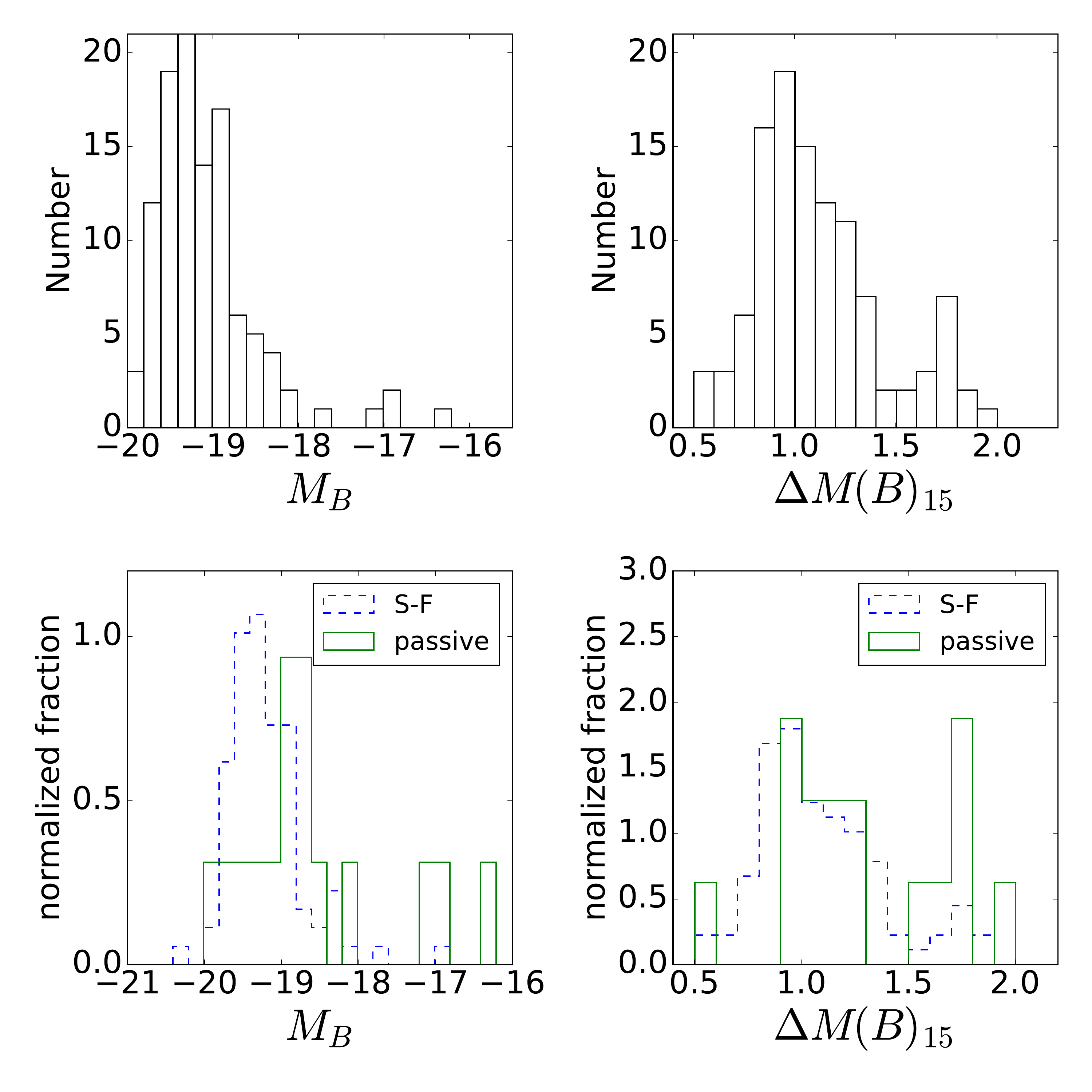}
\caption{Luminosity distribution of observed SNe~Ia in $M(B)$ and $\Delta m_{15}(B)$. The contribution of different host galaxy types is shown. Values are corrected for extinction. The overlaid blue histograms are the distributions of SNe from passive (not star forming) galaxies.}
\end{figure}

Another point that argues for the WD mass being at or near the Chandrasekhar limit (for many normal SNe~Ia) is the following: The ionization balance in late-time spectra of SNe~Ia seems to require neutron-rich Fe-group species. For these elements to be produced, the highest densities that can be achieved in WDs are a necessary condition. Such densities, in turn, are expected only in WDs that are close to the Chandrasekhar mass (e.g., Iwamoto et~al.\ 1999). Furthermore, it appears that the abundance of manganese in the solar neighborhood can only be obtained if a large fraction of SNe~Ia are produced by exploding Chandrasekhar mass WDs (Seitenzahl et~al.\ 2013). This is a consequence of the fact that in order to synthesize manganese (through the production of $^{55}$Co that decays into $^{55}$Fe, which further decays into $^{55}$Mn) densities in excess of $2\times10^8$~g~cm$^{-3}$ are required.

Finally, as we have noted earlier, observations of SN~2014J in the galaxy M82 (at a distance of about 3.5~Mpc) detected clear $^{56}$Co lines at energies of 847 and 1,238 kiloelectron volts (KeV) and a $\gamma$-ray continuum in the 200--400 KeV band (Churazov et~al. 2014). These observations are extremely important for two reasons: (i) they confirm the presence of the decay chain from $^{56}$Ni to $^{56}$Co to $^{56}$Fe that is believed to power (after reprocessing in the expanding ejecta) the optical emission (evidence for the radioactive decay chain was already presented by Kuchner et~al.\ 1994). (ii) The derived mass of radioactive $^{56}$Ni, of about 0.6~$M_\odot$, and the inferred ejecta mass of $\sim$1.2~$M_\odot$ (albeit with a considerable uncertainty), are consistent with a WD mass near the Chandrasekhar value. A sub-Chandrasekhar mass model  would have produced a far too low $\gamma$-ray flux (H\"oflich \& Khokhlov 1996). 

There definitely are, however, also indications that perhaps not all ``normal'' SNe~Ia involve a Chadrasekhar mass WD. In particular, Scalzo, Ruiter \& Sim (2014) used a scaling  relation (which still involved some freedom in the choice of the zero point) between the ejected mass and the light curve width to derive ejecta masses, $M_{ej}$, for 337 SNe~Ia. They also used a relation between the inferred $^{56}$Ni masses, $M_{Ni}$, and the peak luminosity, to derive values of $M_{Ni}$. Reassuringly, the results showed that \textit{at least} 50\% of normal SNe~Ia explode when the WD is at (or near) the Chandrasekhar mass. On the other hand, the results also suggested that as many as 25--50\% of normal SNe~Ia involve sub-Chandrasekhar-mass WDs, with super-Chandrasekhar-mass explosions making up no more than about 1\% of normal SNe~Ia (we shall discuss sub-Chandrasekhar and super-Chandrasekhar models in Section~\ref{sec4}). We should note, nonetheless, that the method used in the Scalzo et~al.\ analysis seems to be sensitive only to the opaque mass. As the Fe-group rich ejecta dominates the opacity, this approach may recover only the Fe-group mass. It would therefore not be surprising for Scalzo et al.\ to find masses that are proportional to the Ni mass.

Considering the uncertainties in all of the above arguments, the next step forward in our attempt to understand SNe~Ia is somewhat less rigorous than we would have liked. Namely, it appears that the only statement that we can make at this stage with some confidence is that: \textit{A significant fraction of normal SNe~Ia are produced by exploding CO white dwarfs when those reach the Chandrasekhar limit.} We cannot exclude the possibility, however, that even ``normal'' SNe~Ia involve some continuous range of masses. In the following discussion of the progenitor systems we shall mainly concentrate on those ``normal'' SNe~Ia resulting from exploding Chandrasekhar-mass CO WDs.

\section{Potential Progenitors\label{sec3}}

Since we have already established that the progenitor system of a SN~Ia has to involve an exploding WD, what we now have to determine is whether this system consists of: (i)~a single, isolated WD, (ii)~a WD and a normal (non-degenerate) companion, (iii)~a WD and a companion that is only the core of a normal star, (iv)~two WDs, or (v)~two WDs and a third star in a triple system. Every single one of these possibilities has actually been suggested as a potential model for the progenitor system! Most of these physical routes may be realistic and indeed may occur in nature. Given the diversity of the observed events it seems that it is important to identify connections between proposed mechanisms and observed SNe, while it seems less likely that a single mechanism is responsible for all SNe~Ia, as many partisans of the various mechanisms often advocate.

In the isolated WD scenario, a single WD explodes as a result of a thermonuclear runaway initiated by pycno-nuclear (density controlled) reactions. In the Single-Degenerate (SD) scenario, the WD grows in mass through accretion from a non-degenerate companion. In the Core-Degenerate (CD) scheme, the WD plunges into the envelope of a giant companion and merges with the giant's core. In the Double-Degenerate (DD) scenario, two white dwarfs in a binary system merge either  relatively smoothly or violently, after having been brought together via the emission of gravitational radiation. Finally, in the triple-system scenario, two WDs are driven to collide through their interaction with a third star in the triple stellar system.

We shall now describe in more detail each one of these scenarios, and discuss what we regard as its main strengths and weaknesses.

\subsection{A Single, Isolated White Dwarf\label{sec3.1}}

The possibility that an isolated CO white dwarf may produce a SN~Ia was recently re-raised by Chiosi et~al.\ 2015 (see also Iben \& Renzini 1983). The idea is that a thermonuclear runaway would be triggered by pycno-nuclear reactions (nuclear reactions caused by high density) between carbon and light elements (including hydrogen and helium), the latter having been internal left-over impurities in the WD. Chiosi et~al.\ suggested that traces of hydrogen and helium would remain inactive in the WD interior, until the WD transits from the liquid to the solid phase. 

While we find this idea interesting and perhaps worthy of further investigation, we do not believe that this can currently be regarded as a viable progenitor model for most normal SNe~Ia, for two main reasons: (a)~It is not clear at all (at least to these authors) whether even the small traces of interior hydrogen and helium required for this model to work, can indeed be produced through normal stellar evolution. In fact, acknowledging this problem, Chiosi et~al.\ (2015) chose to treat the remnant hydrogen and helium mass fractions as free parameters. We should also note that the fact that many WDs are observed to be below the crystallization temperature ($\sim$$3\times10^6$~K for a typical WD) argues that hydrogen is completely depleted in WD interiors. (b) The isolated WD progenitor model typically would lead to explosion (if it indeed works at all) at a mass lower than the Chandrasekhar mass (Chiosi et~al.\ 2015). It therefore cannot explain what appears to be perhaps the main channel of SNe~Ia. In view of these serious difficulties and in the absence of any convincing evolutionary simulations, we shall not discuss this model any further here.

\subsection{The Single-Degenerate Scenario\label{sec3.2}}

In the Single-Degenerate (SD) scenario, a WD in a binary system accretes mass from a normal (nondegenerate) stellar companion, until it reaches the Chandrasekhar mass, at which point explosion occurs (e.g., Whelan \& Iben 1973; Nomoto, Thielemann \& Yokoi 1984).

The companion star itself can, in principle, be a main-sequence star, a subgiant, a red giant (RG), an asymptotic giant-branch (AGB) star, or a helium star (which may have resulted from a star that had lost its hydrogen envelope). The mass transfer from the normal star onto the WD can occur through Roche-lobe overflow (the star filling the critical surface at which tidal forces from the WD overpower the star's own gravity), or through the WD accreting from the powerful stellar wind emitted by the normal star (when the latter is a blue giant, an RG or an AGB star).

Accordingly, the progenitor systems could (again, in principle) include:\break (i)~classical or recurrent novae, or even long-period dwarf novae (in all of which the companion is a low-mass main sequence or a subgiant star; see, e.g., Warner 2003; King, Rolfe \& Schenker 2003), (ii)~systems in which the companion is a young $\sim$6--8~$M_\odot$ (blue) main sequence star, (iii)~steady supersoft x-ray sources (in which the companion is an evolved or subgiant star; e.g., van den Heuvel et~al.\ 1992; Livio 1996), (iv)~symbiotic systems (in which the companion is a red giant or an asymptotic giant-branch [AGB] star; e.g., Webbink et~al.\ 1987), (v)~WD-Hot subdwarf systems (in which the companion is a helium star).

The fact that the SD scenario can potentially involve many progenitor channels, representing binary systems that are all known to exist, may be regarded as a strength, even though the relatively high degree of homogeneity of normal SNe~Ia argues perhaps for one channel dominating over others (if this is indeed the correct scenario). A key question that arises though, is the following: if indeed a large fraction of the normal SNe~Ia represents CO WDs that explode at the Chandrasekhar mass, can the WDs in these potential progenitor systems retain (at least some fraction of) the accreted mass and grow in mass up to the Chandrasekhar limit? (We shall discuss separately, in Section~\ref{sec4}, models in which the WD may explode at a sub-Chandrasekhar or super-Chandrasekhar mass.) The reason that this question is extremely relevant is that there exists substantial observational evidence that at least in some classical nova systems (where the accretion rate is relatively low $\lesssim10^{-8}~M_\odot$~yr$^{-1}$), the WD appears to be losing more mass during nova outbursts than the mass it accretes between outbursts (e.g., Livio \& Truran 1992; Yaron et~al.\ 2005). Similarly, if the accretion rate is too high ($\gtrsim10^{-6}~M_\odot$~yr$^{-1}$), the WD rapidly expands into a red giant configuration (e.g., Nomoto, Nariai \& Sugimoto 1979; Wolf et~al.\ 2013), and the system experiences a common envelope phase (in which the WD and the companion star revolve inside a common envelope that is eventually ejected), which does not lead to an SN~Ia explosion. Consequently, the theoretical consensus has generally been that for the WD to actually grow in mass while accreting hydrogen-rich material, it needs to accrete at a rate in the range that will result in \textit{stable burning}---where the rate at which hydrogen is transformed into helium is equal to the accretion rate (e.g., Paczynski \& Zytkow 1978; Fujimoto 1982; Livio, Prialnik \& Regev 1989; Nomoto et~al.\ 2007; Shen \& Bildsten 2007; Wolf et~al.\ 2013; although see Starrfield 2015 for a somewhat different view), or is at least fairly close to that rate. We should note that to avoid the `straitjacket' imposed by steady burning, it has been proposed that a strong wind emitted by the accreting WD could steer clear of the common envelope phase even at accretion rates higher than those producing steady burning (Hachisu, Kato, \& Nomoto 1996).

Simulations show that steady burning is obtained for accretion rates in the relatively narrow range of (a few) $\times 10^{-8}~M_\odot$ yr$^{-1}$ to (a few) $\times 10^{-7}~M_\odot$ yr$^{-1}$ (e.g., Nomoto et~al.\ 2007; Wolf et~al.\ 2013; Hillman et~al.\ 2016). During the steady burning phase, the WD manifests itself as a supersoft x-ray source (with an effective temperature of 30--100~eV).

Even stable burning of hydrogen, however, does not guarantee a substantial increase in the WD mass, since as helium accumulates, it could lead to helium shell flashes and concomitantly to mass loss (e.g., Newsham et~al.\ 2014; Idan, Shaviv \& Shaviv 2013). A recent simulation that followed the accretion of hydrogen through the steady burning phase and many helium flashes, however, found that the WD can reach the Chandrasekhar mass for a relatively large swath of parameter space (Hillman et~al.\ 2016). The main reason for this somewhat surprising result was that it was found that helium flashes continuously heated the WD. As a result, following a number of explosive flashes, electron degeneracy was significantly lowered in the WD's outer layers, yielding quasi-stable helium burning with no mass ejection. If confirmed by independent simulations, these results could indicate that WDs that accrete hydrogen (or helium) from a normal (or helium) companion at a rate that produces (quasi) steady burning, could actually grow to the Chandrasekhar mass. 

Even this, however, would not necessarily mean that the bulk of normal SNe~Ia are produced by the SD scenario. We need to examine the full range of observational consequences predicted by this model and to compare those with detailed observations. In addition, we need to check whether the \textit{rate} of SNe~Ia events predicted through this route (assuming it can indeed produce SNe~Ia) agrees with the observed rate.

\subsubsection{Strengths of the Single-Degenerate Scenario\label{3.2.1}}

The key strengths of the SD model to SNe~Ia progenitors are the following:
\begin{enumerate}[(1)]
	\item If indeed the accreted mass (or some fraction thereof) can be retained, then the SD scenario provides for a natural, clear path for the WD to reach the Chandrasekhar mass and explode.
	
	\item There exist several specific potential types of progenitor systems, in which the WD mass currently appears to be not too far from the Chandrasekhar mass, and where the accretion rate is sufficiently high so that it may lead to quasi-stable burning or mass growth. Examples include  recurrent novae---systems that have been observed to undergo nova outbursts more than once. The recurrence time of the outbursts scales roughly like $\tau_\mathrm{rec}\sim \frac{R^4_\mathrm{WD}}{M_\mathrm{WD}\dot{M}}$, where $M_\mathrm{WD}$, $R_\mathrm{WD}$ are the WD's mass and radius, respectively, and $\dot {M}$ is the accretion rate (e.g., Truran \& Livio 1986). Since $R_\mathrm{WD}$ decreases with increasing $M_\mathrm{WD}$, recurrent novae that exhibit very short recurrence times (a few years to a few decades) are expected to be characterized by massive WDs and high accretion rates. The mass of the WD in the system U~Sco, for instance, has formally been  determined to be $1.55\pm0.24~M_\odot$ (Thoroughgood et~al.\ 2001), and that in CI~Aql $1.00\pm0.14~M_\odot$ (Sahman et~al.\ 2013).
	
	\item[] Unfortunately, observations do not provide a clear picture as to whether the WD mass is increasing or decreasing in recurrent nova systems. For instance, in the recurrent nova T~Pyx, Selvelli  et~al.\ (2008) and Patterson, Oksanen \& Monard (2013) suggested that the WD is decreasing in mass. However, a more recent analysis by Godon et~al.\ (2014; based on a newly derived distance estimate) concluded that the WD's mass is increasing. Similarly, Schaefer (2013) concluded (with considerable uncertainty) that more mass was ejected during the last outburst of U~Sco than had been accreted prior to the outburst. Still, the fact that the WD mass in recurrent nova systems is high \textit{ab~initio} makes these systems conceivable candidate progenitors of SNe~Ia.  Similar arguments apply to a few nova-like systems such as V~Sge, where a relatively massive WD is thought to be accreting at a high rate from a companion with a mass of $\sim$3~$M_\odot$. Incidentally, in some recurrent novae and symbiotic systems (e.g., T~CrB, RS~Oph) the WD accretes from the wind of a giant companion, and not through Roche lobe overflow (e.g., Munari, Dallaporta \& Cherini 2016; Alexander et~al.\ 2011).	

	\item[] Another class of objects that could, in principle, contain  progenitors, is the group of persistent supersoft x-ray sources. Observations of these sources confirm that they consist of WDs accreting from subgiant companions, at rates that can produce stable burning, i.e., (a few) $\times10^{-7}~M_\odot$ yr$^{-1}$ (e.g., Southwell et~al.\ 1996; Podsiadlowski 2010; and references therein). 
 
	\item[] Finally, in the case of AM~CVn binaries (and the helium nova V445~Pup), the WD is accreting from a helium companion, again making these systems potential progenitor candidates.

\item In the case of SN~2012cg, observations showed some tentative evidence, in the form of excess blue light at fifteen and sixteen days before maximum $B$-band brightness, for the impact of the explosion on a non-degenerate binary companion (Marion et~al.\ 2016). A comparison of these very early data with models by Kasen (2010; see also Marietta, Burrows \& Fryxell 2000; Cao et~al.\ 2015), favored a 6~$M_\odot$ main-sequence companion. However, considering the fact that a binary system consisting of a 6~$M_{\odot}$ main-sequence star and a Chandrasekhar mass WD is rather unlikely, Boehner, Plewa and Langer (2017) concluded (based on their simulations of the impact of the ejecta on the companion) that the most likely companion is a post-main-sequence star, possibly transiting to become a red giant. These observations, while far from being conclusive, are therefore consistent with expectations from the SD scenario (see, however, weakness (i) in Section~\ref{sec3.2.2}, and note that Levanon \& Soker 2017 have shown that blue and UV excess emission could also result from the interaction of the ejecta with matter blown off by an accretion disk in the DD scenario). We should also note that observations suggesting the presence of a companion exist for supernova iPTF14atg (in the form of a UV burst within four days of the explosion; Cao et~al.\ 2015; Liu, Moriya, \& Stancliffe 2015), SN~2008ha (which may, however, have been a core-collapse supernova; Foley et~al.\ 2014; Valenti et~al.\ 2009), and SN~2012z (McCully et~al.\ 2014). However, those were subluminous supernovae with low velocities, whose nature is not entirely clear (e.g., Valenti et~al.\ 2009).

\item The detection of narrow, blue-shifted, Na\,{\small I}~D absorption lines in some SNe~Ia has been taken as evidence of circumstellar material surrounding the progenitor system (e.g., Maguire et~al.\ 2013; Sternberg et~al.\ 2011). In the case of SN~2006X Patat et~al.\ (2007) observed the Na\,{\small I}~D lines to be time-varying. While the origin of this circumstellar material is unclear, it has been argued that one simple explanation for its presence is that it is the result of mass loss from the WD's normal-star companion in the progenitor system. We should note though that potential sources for the origin of the circumstellar material have been suggested also in the context of the double-degenerate scenario (e.g., Shen, Guillochon \& Foley 2013; Raskin \& Kasen 2013), and in particular in the core-degenerate scenario (Soker 2013; see Section~\ref{sec3.3}). We therefore regard the support that the detection of the Na\,{\small I}~D lines provides to the SD scenario as only marginal.
 
	 \item In the case of the well-studied, relatively nearby SN~2011fe, it was found that the fastest moving ejecta (at almost 20,000 km~s$^{-1}$) were composed almost exclusively of carbon (Mazzali et~al.\ 2014). These findings have been interpreted as supporting a scenario in which the WD had been accreting hydrogen-rich material prior to the explosion, and that hydrogen had fused to carbon. In contrast, the merger of two CO WDs would have been expected to produce a considerable amount of oxygen in the outer layers, which was not detected. Again we should note, however, that a different potential explanation for the pure carbon in the fastest moving ejecta was suggested in the context of the core-degenerate scenario (Soker, Garcia-Berro \& Althaus 2014, and see Section~\ref{sec3.3}). Furthermore, in Section~\ref{sec3.2.2} we shall show that SN~2011fe actually contains some of the strongest evidence \textit{against} the SD scenario (effectively ruling out the presence of a giant companion).

	\item In its original formulation, the SD scenario predicted delay times (between the formation of the progenitor systems and the explosion) in a rather narrow range of (a few) $\times10^8-2\times10^9$ years (e.g.\ Han \& Podsiadlowski 2004). This was basically a consequence of the fact that only a narrow range of donor star masses could transfer mass at the rate that could result in stable burning [(a few) $\times10^{-7}\ M_\odot$ yr$^{-1}$]. This prediction was, however, in conflict with observational determinations of delay times that covered a range of about $10^8$--$10^{10}$ years. To overcome this problem and also to increase the predicted SNe~Ia rates, Hachisu, Kato \& Nomoto (1996, 2008) proposed a few modifications to the original SD scenario. First, they suggested that in a system in which the WD accretes mass from a lobe-filling, low-mass red giant, the accreting WD gives rise to an optically thick wind. The mass loss in the wind could stabilize the mass transfer process (since mass loss increases the orbital separation), which would have otherwise been unstable, and also perhaps avoid the formation of a common envelope at high accretion rates (which would have resulted from the fact that red giants that have a convective envelope tend to expand in response to mass loss).
\end{enumerate}

Second, Hachisu et al.\ (2008 and references therein) assumed that the optically thick wind from the WD strips off mass from the outer layers of the donor star, thereby stabilizing the mass transfer process up to main-sequence donors of 8~$M_\odot$. With these modifications to the standard SD scenario, the range of time delays has been significantly extended, with the systems containing massive main-sequence stars providing short delays, and those having red giant donors providing (at least in principle, but see the discussion of weaknesses in Section~\ref{sec3.2.2}) longer delay times.

Thirdly, modeling of SN light curves and late-time spectra (Mazzali et~al.\ 2007) suggests the presence of significant amounts of stable nuclear statistical equilibrium (NSE) Fe-group elements (such as $^{54}$Fe, $^{58}$Ni). These elements are not radioactive but they contribute to the opacity via their large array of spectral lines (Mazzali et~al.\ 2001). They are characterized by a higher neutron number than $^{56}$Ni, and are produced at high densities. Chandrasekhar-mass progenitors reach the required high densities, while WDs with smaller masses do not, thus providing perhaps indirect support for the SD scenario.

Another potential strength for the SD scenario is suggested by the so-called high-velocity features (HVF). These are high velocity ($v>20,000$~km~s$^{-1}$) components to the P-Cygni absorption of the strongest line in SNe~Ia, the Ca~{\sc ii} IR triplet, which are observed at very early times in essentially all SNe~Ia with early enough data (e.g., Mazalli et~al.\ 2005). If the HVF in Ca are very strong, they are also observed in the second strongest line Si~{\sc ii} 6335. Unlike the ``photospheric'' component, HVFs do not evolve in wavelength to lower velocities with time, but tend to remain constant and just to become weaker. This seems to indicate that they are caused by a shell-like distribution of material (clumps or a torus; Tanaka et~al.\ 2008). The high velocity of this material suggests that it is part of the explosion, but normal models do not have enough mass for these lines to form unless the abundances of Ca and Si are unreasonably high just at the outermost layers. Only delayed detonation models place sufficient mass at the velocities of HVFs, but this is still not sufficient. The ionization at low densities is typically higher than what is needed to get strong Ca~{\sc ii} and Si~{\sc ii}. One possible solution is that a higher electron density (than predicted by models) at the highest velocities favors recombination, and a way to achieve this is to add small amounts of H in the mixture (Tanaka et~al.\ 2008). H~masses on the order of a few $10^{-2}~M_{\odot}$ are sufficient and those would not give rise to H$\alpha$ absorption. If this is indeed the correct solution, the~H could be either a remnant of accretion on the surface of the WD or be CSM swept up by the ejecta. Either way, this particular (admittedly speculative) solution would favor the SD channel.

Finally, the lack of polarization in most SNe~Ia (e.g., Wang \& Wheeler 2008) is also a feature that SD models are more likely to reproduce.

Given the clear strengths of the SD Channel, it is difficult to see, at least at the outset, why the single-degenerate scenario would not be the definitive answer for the progenitors puzzle. To understand the reservations, we have to examine the weaknesses of this scenario.

\subsubsection{Weaknesses of the Single-Degenerate Scenario\label{sec3.2.2}}

In its \textit{simplest} form, the SD scenario predicts that at the time of the WD explosion, or at least during an extended priod of time before the explosion, the WD is still in a binary system with a normal (main-sequence, subgiant, giant, or helium-star) companion. The question is then: does the presence of the companion star itself have some specific observational consequences? At least three potential origins for observational signatures come immediately to mind: (i)~The SN ejecta could collide with the companion star. (ii)~Hydrogen could be stripped from the companion and show up in the spectrum. (iii)~If the companion is a giant, its wind would have generated a circumstellar medium, with which the ejecta could interact.

In the following, we examine the observational evidence in relation to each one of these signatures and we show that each one of them can be viewed as a weakness for the SD scenario. 

\begin{enumerate}[(i)]
	\item Concerning the first possibility, the observational consequences of the ejecta interacting with a companion for the early broadband light curve have been calculated in some detail by Kasen (2010), for a red giant, a $6~M_\odot$ main sequence and a $2~M_\odot$ main-sequence companion. Calculation of the impact for main sequence, subgiant, and red giant companions were also done more recently by Boehner, Plewa, and Langer (2017). All the simulations predict optical/UV emission which exceeds the radioactive-decay-powered luminosity of the SN for the first few days following the explosion (although Boehner et~al.\ obtain an energy budget available for prompt emission that is smaller by a factor of 2--4 from that of Kasen). On the observational side on the other hand, three SNe~Ia observed by the Kepler observatory from pre-explosion, with a time resolution of 30~minutes, \textit{show absolutely no signature of ejecta interaction} as predicted by the above models (Olling et~al.\ 2015), in clear conflict with the SD scenario (at least in its basic formulation).
	
	\item[] It has also been suggested that the presence of a companion might leave a signature on the supernova remnant (e.g., Papish et~al.\ 2015). In most SNRs there are no such signatures.
	
\item Clearly the detection of hydrogen in the spectrum is a key diagnostic, since only the SD and the core-degenerate scenarios are capable of exhibiting hydrogen at all in the spectra (either from the stripped normal companion to the WD or from the envelope surrounding the core with which the WD is colliding). In particular, hydrogen stripped from the companion star is predicted to move relatively slowly (with velocities of $\sim$1000~km~s$^{-1}$) and to show up after the higher-velocity outer layers have become optically thin (e.g., Meng, Chen \& Han 2007; Marietta et~al.\ 2000; Liu et~al.\ 2012). Searches for hydrogen in a few normal SNe~Ia have not detected any, with an upper limit on the hydrogen mass of 0.01--0.03~$M_\odot$ (e.g., Leonard 2007; Lundqvist et~al.\ 2013). Most recently, Boty\'anszki, Kasen and Plewa (2018) used multidimentional radiative transfer calculations of late-time spectra with stripped material mixed into the ejecta. They effectively ruled out the presence of main sequence, subgiant, or red giant companions, if the explosion produces $\gtrsim0.1~M_{\odot}$ of unbound solar-abundance ejecta. They used their results to place a strong limit on the hydrogen mass for SN~2011fe [see (iv) below].

\item[] We should note that while a few SNe classified as Ia have shown variable H$\alpha$ line emission (e.g., SN~2002ic; Hamuy et~al.\ 2003 and PTF11kx; Dilday et~al.\ 2012), the relatively high circumstellar mass estimated in the case of PTF11kx argues more for a model in which the SN ejecta interact with the common envelope ejected earlier, as the WD spiraled into the envelope of a giant before colliding with the giant's core (Livio \& Riess 2003; Soker 2013; and see the core-degenerate scenario in Section~\ref{sec3.3}).

\item Similarly, the interaction between a supernova blast wave and circumbinary material is expected to accelerate particles, amplify magnetic fields, and produce radio synchrotron emission (e.g., Chevalier 1998). Radio observations of 27 SNe~Ia with the Very Large Array covering a period of more than two decades have not detected any SN~Ia (e.g., Panagia et~al.\ 2000; Russell \& Immler 2012; Chomiuk et~al.\ 2016). These observations imply a very low density for any circumstellar material, and in turn, impose an upper limit on the mass-loss rate from a putative normal companion of about $5\times10^{-9}\,M_\odot$ yr$^{-1}$. This appears to rule out any symbiotic-type progenitor system, in which the WD accretes from the wind of a red giant. These observations in themselves do not rule out an SD progenitor system in which the WD accretes from a main sequence (or slightly evolved) star, through Roche lobe overflow.

\item[]Other weaknesses of the SD scenario are revealed when one considers observations of particular supernovae.

\item The multiwavelength observations of two relatively nearby normal supernovae, SN~2011fe in the well-studied galaxy M~101 (at a distance of $\sim$7~Mpc), and SN~2014J in M~82 (at a distance of $\sim$3.5~Mpc), deserve special attention (e.g., Chomiuk 2013; Perez-Torres et~al.\ 2014). The wealth of data on these two supernovae place some of the most stringent constraints to date on progenitor models, and they all weaken the case for the SD scenario. We note the following points concerning SN~2011fe:
\begin{enumerate}[(1)]
	\item Deep \textit{pre-explosion} Hubble Space Telescope imaging of the site of SN~2011fe didn't find any source at that location (Li et~al.\ 2011). This rules out the presence of any red giant companion, but cannot exclude subgiant or main-sequence companions with a mass smaller than about 4~$M_\odot$.
	\item No evidence was found for an interaction of the ejecta with a companion (Brown et~al.\ 2012), again ruling out a red giant companion.
	\item Searches for H$\alpha$ emission placed an upper limit of $\sim$0.001~$M_\odot$ (albeit with considerable uncertainty) on the amount of hydrogen that could have been stripped from the companion (Shappee et~al.\ 2013). This limit was made even stronger, $M_\mathrm{stripped} \lesssim10^{-4}~M_{\odot}$, by Boty\'anszki et~al.\ (2018). Similarly, no signs of $P_\beta$ have been reported. Models suggest that $P_\beta$ could be detected as early as one month after the explosion (e.g., Maeda, Kutsuna, \& Shigeyama 2014), and H$\alpha$ about six months after the explosion (e.g., Mattila et~al.\ 2005).
	\item Assuming a spherically symmetric wind profile (from a putative companion) and a wind velocity of 100~km~s$^{-1}$, the non-detection of any radio signal from SN~2011fe places an upper limit on the wind mass-loss rate of $\dot{M}_W \lesssim6\times10^{-10}\,M_\odot$ yr$^{-1}$ (Chomiuk et~al.\ 2012). If confirmed (see e.g., Horesh et~al.\ 2012 for caveats), the low density implied by the radio observations would also rule out an optically thick wind from the accreting WD (assumed as part of the SD scenario; Hachisu et~al.\ 1996, and see discussion above).
	\item The relativistic electrons that can produce radio synchrotron emission can also upscatter photons emitted by the explosion to produce inverse Compton x-ray radiation (e.g., Chevalier \& Fransson 2006). Using the null x-ray observations by the Chandra and Swift observatories, Margutti et~al.\ (2012) constrained the progenitor system's mass-loss rate to $\dot{M} \lesssim2\times10^{-9}\,M_\odot$ yr$^{-1}$, again ruling out any giant companion or substantial mass-loss during Roche lobe overflow.
	\end{enumerate}
\end{enumerate}
	
Overall, the observations of SN~2011fe seem to rule out the presence of a giant companion to the WD, and they may even rule out a main sequence companion, other than in variations of the SD scenario which allow the companion to shrink by orders of magnitude in radius (such as the spin-up/spein-down scheme, see below).

It is also worth noting that with a main-sequence (or slightly evolved) companion one might have expected the progenitor system to have undergone in the past recurrent nova outbursts or supersoft x-ray emission episodes. No evidence for a nova outburst or nova shells (ejected during outbursts) was found in pre-explosion x-ray images of the site of SN~2011fe (e.g., Liu et~al.\ 2012; Nielsen, Voss \& Nelemans 2012).

In the case of SN~2014J the observational limits become in some respects, even stronger:
\begin{enumerate}
	\item[(1)] Deep x-ray observations revealed no x-ray emission down to a luminosity of $L_x<7\times10^{36}$~erg~s$^{-1}$ in the range 0.3--10~KeV (Margutti et~al.\ 2014), implying a low-density environment with $n<3$~cm$^{-3}$ and constraining the pre-explosion mass-loss rate to 
	$\dot{M}<10^{-9}~M_\odot$~yr$^{-1}$.
	\item [(2)] Similarly, null radio observations placed an even more stringent limit of $\dot{M}\lesssim7\times10^{-10}~M_\odot$~yr$^{-1}$ (for an assumed wind speed of 100~km~s$^{-1}$; Perez-Torres et~al.\ 2014).
\item[] Taken together, these results rule out SD scenarios for SN~2014J in which the WD accretes from the wind of a giant companion (symbiotic systems). They also exclude much of the parameter space for systems in which the WD accretes through Roche lobe overflow (from a main-sequence, subgiant or helium-star companion) at rates that produce stable burning (a few $\times10^{-7}~M_\odot$~yr$^{-1}$), if at least 1\% of the transferred material is lost from the system.
\item[(3)] Near-UV to near-IR pre-explosion imaging with the Hubble Space Telescope (HST) did not detect any progenitor system at the site of SN~2014J (Kelly et~al.\ 2014), again excluding all SD progenitor systems that include a red-giant companion to the WD. We should note though that the dust  reddening towards SN~2014J is rather substantial.
\item[]Another test for the SD scenario is the existence (or absence) of a companion star that survives the explosion. Searches for surviving companions have been performed in a few supernova remnants (SNRs).
\item[]Observations of one supernova remnant provide especially stringent constraints:
\begin{enumerate}
	\item[(v)] Observations of the central region of the $400\pm50$ years old supernova remnant SNR~0509--67.5 strongly contradict expectations from the SD scenario. These HST observations of the site of an SN~Ia in the Large Magellanic Cloud, revealed no candidate companion down to a limiting absolute magnitude of $M_\mathrm{V}=+8.4$ (Schaefer \& Pagnotta 2012). On the face of it, this absence of an ex-companion rules out most (if not all) SD models in the case of the SN~Ia that created SNR 0509--67.5.
	\item[]In the case of SN~1006, the non-detection of a companion excludes the possibility of a red giant (Gonz\'alez Hern\'andez et~al.\ 2012).
	\item[]The situation concerning the SNR of Tycho's 1572 supernova is somewhat more ambiguous. While a type G0--G2 star has been suggested to potentially be the surviving companion (based on its velocity and composition; Ruiz-Lapuente et~al.\ 2004; Bedin et~al.\ 2014), this identification has been put into question by other studies (e.g., Kerzendorf et~al.\ 2013).
	\item[]In an attempt to reconcile these non-detections with the SD scenario, some researchers rely on a `spin-up/down' mechanism (e.g., Justham 2011; Di~Stefano, Voss \& Claeys 2011; Hachisu, Kato \& Nomoto 2012, Boshkayev et~al.\ 2013; Wang et~al.\ 2014). In this scenario, following the accretion phase, the WD is spinning at such a high rate that the rotational support can prevent it from exploding, even at a super-Chandrasekhar mass. Carbon burning is initiated only after the WD loses its angular momentum through magnetic braking. In this scenario, the spin-down period is sufficiently long (e.g., $>$$10^5$~yr; Di~Stefano \& Kilic 2012), so that the RG companion has time to evolve to an undetectable He white dwarf before the explosion, thus explaining the absence of a companion. Attempts to empirically constrain the spin-down timescale concluded that it may be too early to rule out the possible existence of a dim, undetectable companion in SNR~0509-67.5 (Meng \& Podsiodlowski 2013).
	\item[(vi)] In the SD scenario, the WD is expected to accrete at a high rate ($\gtrsim10^{-7}~M_\odot$~yr$^{-1}$) and burn hydrogen steadily for a relatively long period of time prior to the explosion. During at least part of this phase, the WD can be a supersoft x-ray source with an effective temperature (defined by the Stefan-Soltzmann law) of $T_\mathrm{eff}\simeq45$ ($\dot{M}/10^{-7}~M_\odot$ yr$^{-1})^{1/4}(R_\mathrm{WD}/10^{-2}~R_\odot)^{1/2}$ eV. Given the known rate of SNe~Ia (one in a few hundred years for a typical galaxy), one can estimate that at any given time a galaxy should contain (a few)$\times(10^2$--10$^3)$ such accreting WDs. Gilfanov \& Bogd\'an (2010) compared the predicted x-ray luminosity (assuming it persists) to the observed one for six nearby elliptical galaxies and galaxy bulges, and found the observed integrated x-ray flux to be 30--50 times lower than predicted. They tentatively concluded that the SD scenario can be responsible for no more than about 5\% of the SNe~Ia in elliptical galaxies.
	\item[] Earlier work by Di Stefano (2010) also showed that the actual \textit{numbers} of supersoft x-ray sources in six galaxies were smaller than those expected if the simple version of the SD scenario were to account for the observed SN~Ia rate (this does not necessarily mean that the SD scenario does not produce SNe~Ia at all, since this channel may be less expected to occur in ellipticals). 
	\item[] Hachisu, Kato \& Nomoto (2010) argued that due to the optically thick wind from the accreting WD, the supersoft sources are hidden from view during the high-accretion-rate phase, with the x-ray radiation being reprocessed into UV emission. However, searches for such unusual UV-bright objects in the Small Magellanic Cloud have also failed to find any (Lepo \& van Kerkwijk 2013) and searches for the ionized helium and forbidden emission lines of C, N, O, expected to result from the ionizing radiation of such UV sources have also not found any evidence for their existence (e.g., Woods \& Gilfanov 2013, 2014; Johansson et~al.\ 2014). These results also seem to place an upper limit of about 10\% on the potential contribution of the standard SD scenario to the SNe~Ia rate.
	\item[(vii)] Even with the modifications to the original SD scenario that include an optically thick wind from the accreting WD, and mass stripping (by that wind) from the outer layers of the donor star, the SD scenario still seems to fail to produce an SNe~Ia rate (per unit mass) that agrees with observations at long delay times (e.g., Bours, Toonen \& Nelemans 2013). The main reason for this discrepancy is that theories of binary star evolution do not produce many binary systems containing a red giant and a white dwarf with orbital periods of about a year or more (e.g., Yungelson \& Livio 2000). This paucity, in turn, is a consequence of the fact that binaries containing a WD are typically formed via a common-envelope phase. In this process the secondary star in the binary system is engulfed inside the envelope of a giant (or AGB star), resulting in the red giant's core (a WD in the making) and the secondary spiraling in, thereby eventually ejecting the common envelope.
	\item[] As a result, binary systems that emerge from a common envelope tend to have shorter orbital periods than a few hundred days.
	\item[] Consequently, without further modifications, the SD scenario fails to produce the $t^{-1}$ behavior of the DTD over the entire range of delay times at which the DTD is observed. Again, this may be fatal for the SD scenario \textit{only if we insist on one type of progenitor model for all SNe~Ia}.
	\end{enumerate}
\end{enumerate}
	
\subsection{The Core-Degenerate Scenario\label{sec3.3}}

The core-degenerate (CD) scenario for SNe~Ia was originally suggested by Livio \& Riess (2003) to specifically address the detection of a broad H$\alpha$ emission component in the spectrum of the SN~Ia SN~2002ic (Hamuy et~al.\ 2003). This detection was puzzling because H$\alpha$ of the same strength could have easily been detected (were hydrogen present) in at least 100 previous SN~Ia spectra, but it hadn't been. Consequently, Livio and Rees proposed the following scenario:

All the binary evolutionary tracks that lead to the formation of close double-WD systems involve a stage at which an asymptotic giant branch (AGB) star fills its Roche lobe and transfers mass onto its WD companion. Since this mass-transfer process is unstable (the AGB star expands upon losing mass because it has a convective envelope), the system rapidly evolves into a common envelope (CE) configuration, in which the WD and the core of the AGB star spiral-in, inside a CE (e.g., Rasio \& Livio 1996; Taam \& Sandqvist 2000). Usually, this process results in the ejection of the CE and the emergence of a close double-WD system (initially the core of the AGB star is a subdwarf, which becomes a WD upon cooling and shrinking). Livio \& Riess (2003) proposed that in the case of SN~2002ic, the WD and the core merged, and the explosion ensued while there was still CE material around, for the SN ejecta to interact with.

This scenario has later been picked up and expanded upon by Soker and collaborators (e.g., Kashi \& Soker 2011; Ilkov \& Soker 2012) to constitute a potential progenitor scenario for SNe~Ia in general. The idea is that the WD merges with the core of the AGB star within less than $\sim$10$^5$~yr after the initiation of the CE phase (potentially still in a planetary nebula phase), while the core (which is sometimes more massive than the WD) is still relatively large and hot.

In the CD scenario, the merged product is a super-Chandrasekhar object, but it doesn't explode because it is supported by rotation (e.g., Yoon \& Langer 2005). Consequently, most of the delay time between the binary formation and the SN explosion in the CD scenario is supposed to be caused by the spin-down time (through a magneto-dipole radiation torque) of the merger product.

We should recall that rotational support as a means to achieve super-Chandra\-sekhar WDs and/or to delay the explosion has also been suggested in the context of other scenarios, such as the SD version (e.g., Di Stefano, Voss \& Claeys 2011; Tornamb\'e \& Pirsonti 2013; Hachisu, Kato \& Nomoto 2012) that involves a spin-up/down mechanism designed to allow the companion star to shrink. The stability (or not) of such differentially rotating configurations is not entirely clear (e.g., Yoon \& Langer 2005; Piro 2008; Hachisu et~al.\ 2012). We should also note that ultramagnetized WDs, which were purported to have a higher maximum mass limit (than the standard Chandrasekhar mass), were later shown to be unstable (Coelho et~al.\ 2014). We shall discuss this issue further when we'll consider super-Chandrasekhar models in Section~\ref{sec4}.

\subsubsection{Strengths of the Core-Degenerate Scenario\label{sec3.3.1}}

The main strengths of the CD scenario are the following:

\begin{enumerate}[(1)]
	\item Since the explosion is expected to occur only after the merged product of the WD and the core of the AGB star has spun down, and since this spin-down period \textit{may} take a long time (in principle at least, although Meng \& Podsiadlowski 2013 argued on empirical grounds that it may be shorter than $10^7$~yr), the fact that neither a companion nor evidence for any circumstellar hydrogen was found in many SNe~Ia can be explained. Material from the common envelope could have long dispersed and the explosion itself involves only one isolated WD. This could also explain the near spherical symmetry of most supernova remnants of SNe~Ia.
	
	\item At the same time, in cases in which the spin-down time is relatively short (assuming that a wide range in spin-down times is possible), the presence of a relatively massive ($\gtrsim$$1~M_\odot$) surrounding shell of circumstellar hydrogen (as in the case of SN~2002ic and PTF~11kx) is also possible. In fact, Tsebrenko \& Soker (2015) have even argued that some SNe~Ia may have exploded inside planetary nebulae (the ejected outer layers of AGB stars that are ionized by radiation from the central hot core). The CD scenario provides a natural mechanism for the subclass of SNe~Ia, known as SNe~Ia-CSM, because they show signs of the SN ejecta ramming into a dense circumstellar medium (see Section~\ref{sec4}). Most recently, even x-ray emission was detected from SN~2012ca (Bochenek et~al.\ 2017).
	
	\item As we have noted earlier, detailed observations of the relatively close SN~2011fe indicate that the fastest moving ejecta are composed almost exclusively of carbon (Mazzali et~al.\ 2014). It has been shown that \textit{if} the delay time is sufficiently long ($\tau_{delay}\gtrsim1.4$~Gyr), the WD crystallizes, with the denser oxygen-rich solid sinking down, and the outer layers becoming rich ($\sim$90\% by mass) in carbon (Soker, Garc\'{\i}a-Berro, \& Althaus 2013). 
	
	\item[] We should note, however, that crystallization may lead to the WD undergoing an accretion-induced-collapse into a neutron star instead of an SN~Ia explosion (e.g., Nomoto \& Kondo 1991).
	
	\item The line-of-sight to highly reddened SNe~Ia shows peculiar continuum polarization, steeply rising toward blue wavelength and peaking at $\lambda\lesssim0.4~\mu$m. Since some proto-planetary nebulae have similar polarization curves, it has been speculated that this indicates an explosion during the post-AGB phase (of the companion), thereby supporting the CD scenario (Cikota et~al.\ 2017).
	
	\end{enumerate}
	
\subsubsection{Weaknesses of the Core-Degenerate Scenario\label{sec3.3.2}}

The principal weaknesses of the CD scenario are the following: 

\begin{enumerate}[(i)]
	\item Perhaps the biggest weakness is the fact that to date there are no detailed simulations of the process of the merger of a WD with the core of an AGB star (for a preliminary simulation see Aznar-Sigu\'an et~al.\ 2015). Consequently, it is actually not clear whether such a merger indeed results in anything that resembles an SN~Ia explosion (including, e.g., the amount of $^{56}$Ni produced).
	
	\item Given the uncertainties in the merger process and the fact that rotation can in principle support a super Chandrasekhar mass, it is not obvious that the explosion typically occurs when the WD has a Chandrasekhar mass (and not at super-Chandrasekhar masses). If this process indeed more naturally produces explosions in WDs that have a super-Chandrasekhar mass, then it cannot account for the majority of SNe~Ia.
	
	\item While the spin-up/spin-down scheme has also been invoked in the context of the SD scenario (e.g., Di Stefano, Voss \& Claeys 2011; Justham 2011; Yoon \& Langer 2005; Hachisu, Kato \& Nomoto 2012) to account for the apparent absence of circumstellar hydrogen in a number of well-studied SNe~Ia, the precise processes involved in the spin-up and spin-down phases are highly uncertain (the induced compression in white dwarfs, caused by the loss of angular momentum, was studied e.g., by Boshkayev et~al.\ 2016). Observations of other systems that involve accreting white dwarfs, such as cataclysmic variables, present a somewhat ambiguous picture as to whether the WDs can truly reach significantly super-Chandrasekhar masses. For example, for spin-up to be achieved in these systems at all, the accreting WD has to avoid nova eruptions and accretion at rates that produce expansion to giant dimensions, otherwise the interaction between the WD and the envelope results in spin-down (e.g., Livio \& Pringle 1998). Similarly, it is not clear whether the WD will be forced to rotate as a rigid body (e.g., through the baroclinic instability, Piro 2008), in which case the maximal mass that can be rotationally supported is only slightly higher than the Chandrasekhar mass, or whether it can rotate differentially (e.g., Yoon \& Langer 2005). As a consequence of these uncertainties, it is not obvious that the CD scenario can indeed work as conceived.
	
	\item There is no known compelling reason why the CD scenario should produce a Delay Time Distribution that is proportional to $t^{-1}$ (or even $t^{-1.5}$, as suggested by Heringer et~al.\ 2017), although any scenario in which the time to explosion is a steeply decreasing function of some parameter should produce such a DTD. 
\end{enumerate}	

\subsection{The Double-Degenerate (DD) Scenario\label{sec3.4}}

\subsubsection{Tidal Mergers Involving an Accretion Phase\label{sec3.4.1}}

In the original double-degenerate scenario, two CO WDs of different masses in a binary system are brought together by the emission of gravitational radiation (Webbink 1984; Iben \& Tutukov 1984). The total mass of the system is assumed to exceed the Chandrasekhar mass. Once the lighter of the two WDs (which has a larger radius) fills its Roche lobe, a dynamically unstable mass-transfer process ensues, and the mass donor is totally tidally disrupted within a few orbital periods (tens of seconds), to form a massive disk around the primary WD (e.g., Rasio \& Shapiro 1994; Pakmor et~al.\ 2012; Shen et~al.\ 2012; Mall et~al.\ 2014).

The subsequent evolution of this configuration depends largely on the precise WD masses and on the accretion rate from the disk onto the primary WD. Given that in this scenario the accretion is of carbon- and oxygen-rich material (thus avoiding altogether nova eruptions and helium shell flashes), it has been shown that at least under some conditions the merger can lead to central carbon ignition and an SN~Ia (e.g., Yoon et~al.\ 2007; Sato et~al.\ 2015).

\subsubsection{Violent Mergers\label{sec3.4.2}}

Since the merger of two WDs is a complex hydrodynamical process, understanding its details requires high-resolution 3D simulations. A few recent such simulations revealed that carbon detonation can occasionally occur due to compressional heating (for example, where an accretion stream hits the surface of the primary WD) during a violent merger phase (e.g., Pakmor et~al.\ 2012; Raskin et~al.\ 2014; Don et~al.\ 2014; Sato et~al.\ 2015). Sato et~al.\ (2015) investigated the range of masses of the merging WDs that can lead to an SN~Ia explosion through such a violent verger, and concluded that those happen when \textit{both} WDs have relatively high masses, in the range $0.9~M_{\odot}\lesssim M_\mathrm{WD}\lesssim1.1~M_\odot$. They also found that for primary masses in the range $0.7~M_{\odot}\lesssim M_{1}\lesssim0.9~M_\odot$, but when the total mass of the two components still exceeds the Chandrasekhar limit, dynamical carbon ignition does not occur during the merger itself. Nevertheless, if carbon is not ignited off-center during the accretion phase, central carbon ignition still occurs, followed by an SN~Ia explosion.	

\subsubsection{Third-Body-Induced Collisions\label{sec3.4.3}}

Since violent collisions can (in principle at least) produce SNe~Ia, it is only natural to also consider head-on collisions that do not result from the spiraling-in of two WDs due to gravitational radiation losses. Such collisions can occur either in unusually dense stellar environments, or as a result of a third body that influences the orbit of a binary WD system. The first situation may arise, for instance, in globular clusters (e.g., Rosswog et~al.\ 2009; Raskin et~al.\ 2010; Lor\'en-Aguilar, Isern \& Garc\'{\i}a-Berro 2010), but it is quite clear that it cannot account for the majority of SNe~Ia, in terms of the expected rates of such collisions.

The idea that a third, main-sequence star, in a hierarchical-triple system containing a binary WD can induce Kozai-Lidov oscillations in the inner binary, driving it to high eccentricity and collision, was first introduced by Thompson (2011). Basically, the two orbits (of the binary WD and of the more distant main-sequence star) torque each other and exchange angular momentum, and for highly inclined triple systems large amplitude oscillations of the eccentricities can be produced. Katz \& Dong (2012) showed that WD--WD binaries with a semi-major axis of $a\sim1$--300~AU orbited by a third body with a pericenter ratio $r_{p,out}/a\sim3$--10, have a chance of a few percent to experience a head-on collision within 5~Gyr. Kushnir et~al.\ (2013) further suggested that this could be the main scenario leading to SNe~Ia.

\subsubsection{Strengths of the Double-Degenerate Scenario\label{sec3.4.4}}

The main strengths of the DD scenario (in its three different variants) are the following:
\begin{enumerate}[(1)]
	\item Double-degenerate systems are a natural product of binary star evolution following one or two common envelope events, and such systems have been detected observationally (e.g., Iben \& Tutukov 1984; Iben \& Livio 1993; Saffer, Livio \& Yungelson 1998; Roelofs et~al.\ 2010).
	\item The merger or collision of two WDs in a close binary system is likely to produce some significant outcome (an explosion, a very compact object, or a giant like the objects known as R~Cr~Br stars).
	\item A merger or collision of two WDs would be generally  consistent with the non-detection of circumstellar hydrogen, the non-detection of a companion star in the relatively close supernovae SN~2011fe and SN~2014J (although see some reservations about WD mergers in the case of SN~2011fe; Levanon, Soker \& Garc\'{\i}a-Berro 2015), and the null detection of a companion in the supernova remnant SNR~0509--67.5 in the LMC.
	\item The behavior of the Delay Time Distribution is obtained naturally for two WDs brought together by the emission of gravitational radiation, at least for a certain range of delay times (e.g., Yungelson \& Livio 2000). The distribution of the separation, $a$, of main-sequence binaries behaves roughly as $d$N$/da\sim a^{-1}$ (e.g., Duquennoy \& Mayor 1991). Populations synthesis simulations of the evolution of binary systems show that the separations of binary WD systems that emerge from common envelope evolution have a similar distribution (e.g., Toonen, Nelemans \& Portegies Zwart 2012). The delay time till the mergers of such double WD systems is determined primarily (except for delays shorter than about 1~Gyr, see below) by the timescale of the orbit's decay due to gravitational radiation, $t\sim a^4$. Consequently, merging WDs have a delay-time-distribution (for delay times longer than 1~Gyr) $d$N$/dt=d$N$/da\ da/dt\sim t^{-1}$, in clear agreement with the observations of the DTD (as we have noted earlier, Heringer et~al.\ 2016 find under some simplifying assumptions $t^{-1.5}$). We should note that in the variant in which a WD--WD direct collision is triggered by a third-body main-sequence perturber in a triple system, the time to collision behaves like $t\sim a^{5/2}$ (Katz \& Dong 2012). This again produces a DTD that is approximately proportional to $t^{-1}$, even if the distribution of the separations of WD binaries in triple systems is not precisely $d$N$/d a\sim a^{-1}$.
	\item[] For short delay times, the delay time is determined primarily by the main-sequence lifetime of the lighter of the two (WD-producing) main-sequence stars, and by the evolution of the initial binary. Consequently, for delay times shorter than $\sim$1~Gyr (corresponding to the lifetime of 2--3~$M_\odot$ stars), the distribution behaves like the production rate of WDs, which is proportional to $t^{-1/2}$ (Pritchet, Howell \& Sullivan 2008). 
	\item Numerical simulations of the head-on collision of two 0.7~$M_\odot$ WDs, induced by a third body perturber, for instance, have shown that a detonation ensues, and that as much as 0.56~$M_\odot$ of $^{56}$Ni can be synthesized, in agreement with observations of typical SNe~Ia (Kushnir et~al.\ 2013). In addition, Dong et~al.\ (2015) have shown that in two (or possibly three) SNe~Ia (out of $\sim$20), double-peaked line profiles have been observed in their nebular spectra. Dong et~al.\ argued that these double-peaked profiles reflect a bi-modality in the velocity distribution of the $^{56}$Ni in the ejecta. Such a bi-modality is naturally expected in a direct collision of two WDs, as the two WDs detonate, and indeed it was demonstrated in the numerical simulations.	
\end{enumerate}

\subsubsection{Weaknesses of the Double-Degenerate Scenario\label{sec3.4.5}}

In spite of the interesting attributes of the various possible channels of the DD scenario, they also have several clear weaknesses.
\begin{enumerate}[(i)]
	\item One significant weakness is theoretical: in what is perhaps its most straightforward variant, the DD scenario appears to be leading to an accretion-induced collapse (AIC), rather than to an SN~Ia. This is the path in which the lighter WD is tidally disrupted to form a disk around the primary, and carbon is ignited at the boundary between the accretion disk and the heavier WD (e.g., Mochkovitch \& Livio 1989, 1990; Saio \& Nomoto 1998). Sato et~al.\ (2015) found that such AICs are obtained for a primary mass in the range 1.0--1.1~$M_\odot$ and a secondary mass in the range 0.5--0.7~$M_\odot$. Comparing their results to the distribution of WD masses (e.g., Kleinman et~al.\ 2013) and the merger rates obtained in population synthesis simulations (Badenes \& Maoz 2012), Sato et~al.\ argued that the DD scenario can account only for $\lesssim9$\% of the Galactic rate of SNe~Ia.
	\item The DD scenario that invokes triple stellar systems to prompt a collision also involves a number of serious uncertainties. First, it is far from clear that a sufficient number of double white dwarf systems will find themselves in triples with the appropriate orbital elements. This is due both to the general frequency of triple stellar systems (which is at the 10--20\% level, e.g., Leigh \& Geller 2013), and to the fact that the inner pair in a hierarchical triple might collide already when the stars are on the main sequence (e.g., Hamers et~al.\ 2013). In fact, a recent study on the statistics of wide-orbit double-degenerate systems has placed a tight constraint on the viability of the WDs collision model as the primary channel for SNe~Ia (Klein \& Katz 2017). The study found that $\sim$10\% of the wide-orbit binary WD systems should end up in a collision to account for the SNe~Ia rate. This is in tension with the estimate that only a few percent of triple systems with wide-orbit double WDs having the right hierarchy lead to a collision (Katz \& Dong 2012). Observations with the Gaia Space Observatory should give a clearer picture as to whether the statistics of triple systems support the DD collision induced by a third star scenario.
	\item[] The precise outcome of triple-induced WD collisions that are not precisely head-on is also not entirely clear, even though preliminary simulations seem to indicate that very similar detonation are obtained for values of $r/(R_1+R_2)$ of up to 0.5 (where $r$ is the impact parameter and $R_1, R_2$ the radii of the two WDs), and for higher values of this ratio no explosion ensues (D.~Kushnir, private communication). 
	\item[] Note also that in violent mergers \textit{any one of the exploding WDs is in fact at a sub-Chandrasekhar mass}. This reflects on the density at which ignition and flame propagation occur. It is not clear, therefore, that the precise velocity distributions observed in normal SNe~Ia can be produced (see discussion in Section~\ref{sec2.2}).
	\item[] In addition, synthesis of neutron-rich Fe group isotopes is very difficult, if not impossible, if $M_{\rm WD}<M_{ch}$. It is not clear if the same difficulty applies also to the third-body-induced-collision scenario, since in the collision process itself there is adiabatic compression and also the density jumps by a factor 4 due to the shock wave. Whether this is sufficient to compensate for the central density being lower than in a Chandrasekhar mass WD will require further investigation (e.g., it turns out that $^{58}$Ni can be produced if one assumes the presence of small amounts of $^{22}$Ne).
	\item Other than the double-peaked line profile mentioned above (in at most~3 out of 20~SNe~Ia; Dong et~al.\ 2015), SNe~Ia generally exhibit a fairly high degree of spherical symmetry. Violent mergers and collisions, on the other hand, produce asymmetrically structured ejecta, which in turn produce variations as a function of viewing angle far and beyond those observed in the spectra and light curves. In particular, Bulla et~al.\ (2016) have shown that a violent merger of two CO white dwarfs has serious difficulties in reproducing the very low polarization levels commonly observed in normal SNe~Ia (e.g., Wang \& Wheeler 2008; highly polarized events, such as SN~2004dt are rate, e.g., Leonard et~al.\ 2005; Wang et~al.\ 2006). In general, the percentage of polarized flux is expected to scale with the projected axis ratio of the source of the photons. Since the continuum polarization is generally not more than a fraction of a percent (e.g., Wang \& Wheeler 2008), a very slight asymmetry is indicated (and even that only in the outer parts of the expanding photosphere). Similarly, a recent survey of polarization measurements concluded that the delayed detonation model for the explosion (Nomoto 1982; Khokhlov 1991) could explain the existing observations (Meng, Zhang, \& Han 2017; see though Cikota et~al.\ 2017). 
	\item[] The morphology of normal SNe~Ia is also generally consistent with spherical symmetry as is the morphology of most SNe~Ia remnants, again in contrast with expectations from violent mergers or collisions. This is the situation, for instance, with the relatively well-observed normal supernovae SN~2011fe and SN~2012fr (e.g., Maund et~al.\ 2013; Soker, Garcia-Berro \& Althaus 2014). We should note though that spectropolarimetry studies of the subluminous SN~2005ke and SN~1999by did indicate deviations from spherical symmetry in these systems (e.g., Patat et~al.\ 2012; Howell et~al.\ 2001).
\end{enumerate}

\section{Additional Subclasses\label{sec4}} 

Over the years, observations showed that in addition to the ``normal'' SNe~Ia there are several subclasses of outliers, as well as a considerably larger diversity than originally realized (see, e.g., Taubenberger 2017). At first, those included supernovae that were somewhat brighter than the norm, of which SN~1991T was a prominent member, and supernovae that were dimmer than the norm, like SN~1991bg. SN~1991T was very luminous, and showed spectra dominated by lines of doubly ionized species (in particular Fe~{\sc iii}) at early times (Filippenko et~al.\ 1992a; Phillips et~al.\ 1992). The high ionization was explained by the larger $^{56}$Ni mass (Ruiz-Lapuento et~al.\ 1993) and by the wider distribution of $^{56}$Ni in velocity (Mazzali, Danziger \& Turatto 1995; Sasdelli et~al.\ 2014). At late times, however, SN~1991T looks similar to normal SNe~Ia, apart from the higher flux. The mass of $^{56}$Ni has been estimated to be close to $1~M_{\odot}$ (Mazzali et~al.\ 2007), indicating very efficient burning of a likely $M_\mathrm{ch}$ progenitor, or possibly a slightly overmassive progenitor. After SN~1991T, other SNe have been found that have properties intermediate between SN1991T and normal SNe~Ia (e.g., SN~1999as; SN~1999ac), and some that are even more extreme in being burned to $^{56}$Ni such as SN~2011hr (Zhang et~al.\ 2016).

At the opposite end of the luminosity range, SN~1991bg showed a rapidly evolving light curve and strong Ti~{\sc ii} lines (Fillipenko 1992b; Leibundgut 1993). This was shown to be mostly the effect of temperature as a consequence of the low luminosity (Nugent et~al.\ 1995; Mazzali et~al.\ 1997). The late-time spectra of SN1991bg, however, behave differently from other SNe~Ia, with a dramatic narrowing of the lines after $\sim$200 days and the surviving of [Fe~{\sc iii}] lines, while [Fe~{\sc ii}] lines fade. This is indicative of a low density in the inner part of the ejecta, suggesting low mass(es) of the progenitor(s), possibly as the result of a WD merger (Mazzali \& Hachinger 2012). 

Later, three other subclasses have been identified. These included a group sometimes referred to as supernovae Type~Iax, overluminous SNe~Ia, and SNe in which there was evidence for interaction with a relatively dense circumstellar medium---SNe~Ia-CSM.

SNe~Iax are fainter than normal SNe~Ia; they involve lower velocities, and have a bluer continuum at early phases. They do not obey the Phillips relation (e.g., Foley et~al.\ 2009; Phillips et~al.\ 2007).

Overluminous SNe~Ia exhibit luminosities that seem to require a $^{56}$Ni mass in excess of $1~M_{\odot}$, which, in turn, seems to correspond to a super-Chandrasekhar-mass white dwarf (e.g., Howell et~al.\ 2006; Scalzo et~al.\ 2010).

The third subclass, SNe~Ia-CSM, exhibit strong Balmer lines in emission, most likely obtained as the SN~ejecta ram into a dense CSM (e.g., Hamuy et~al.\ 2003; Aldering et~al.\ 2006; Silverman et~al.\ 2013; Leloudas et~al.\ 2015). In many respects these supernovae resemble the brighter-than-normal SN~1991T.

\textit{There are no unambiguous progenitor models for any of these subclasses}. However, several lines of evidence point to SNe~Iax potentially being associated with the SD scenario. In particular, a companion may have been identified in a pre-explosion image of the SN~Iax~2012Z (McCully et~al.\ 2014), and post-explosion in the SN~Iax~2008ha (Foley et~al.\ 2014; which may, however, have been a core-collapse supernova; Valenti et~al.\ 2009). Uncertainty still remains, since the putative companion in the case of SN~2012Z is a blue source, while in the case of SN~2008ha the candidate surviving companion is a red point source.

Another piece of evidence potentially suggesting an SD scenario for the progenitors of SNe~Iax is the tentative detection of an infrared light echo (for at least excess of mid-IR emission) from the SN~Iax 2014dt (Fox et~al.\ 2016). An IR light echo is expected when photons from the explosion are re-emitted by dust grains in circumstellar material, and the existence of such material is generally expected in the SD~scenario.

We should note that it has also been suggested that some of the fainter, less energetic events, such as SN~2002cx and SN~2005hk (dubbed SNe~Iax) themselves may be associated with deflagrations that failed to transition into detonations (e.g., Sahu et~al.\ 2008). A weak deflagration may not be able to obliterate the exploding WD, but rather it would leave behind a compact remnant (e.g., Kromer et~al.\ 2013).

As we noted above, a number of SNe that show the characteristic spectrum of an SN~Ia are so luminous that they seem to be incompatible with the explosion of a Chandrasekhar-mass WD. They are therefore generally referred to as super-Chandrasekhar supernovae. Following the first such event (Howell et~al.\ 2006), a few other caes have been found, with SN~2007if (Yuan et~al.\ 2010; Scalzo et~al.\ 2010; Childress et~al.\ 2011) and SN~2009dc (Taubenberger et~al.\ 2011) being among the best observed. They exhibit a hot spectrum, similar to that of SN~1991T, which is in line with a high luminosity and a low velocity. The latter may be the consequence of a very high binding energy of a massive progenitor WD. For super-Chandrasekhar supernovae, models usually rely on the spin-up (during the accretion phase) spin-down (prior to the explosion) scenario, for the WD to achieve a super-Chandrasekhar mass. As we have noted earlier, these models also typically assume that the spin-down phase is sufficiently long, so that any red giant donor (in the SD-scenario) would have had time to evolve to a hard-to-detect He white dwarf before the SN~explosion.

We should note though that spectral models of super-Chandrasekhar supernovae have not been able to reproduce the combination of high luminosity, low velocities and high temperature. An alternative option was therefore presented (e.g., Hachinger et~al.\ 2012), in which much of the flux is not caused by $^{56}$Ni, but rather by the interaction of the SN ejecta with a hydrogen- and helium-poor CSM. A possible scenario might then be the violent merger of two massive WDs.

Finally, the SNe~Ia-CSM require a substantial amount of mass to be lost before the supernova explosion. This is, in fact, most consistent with the core-degenerate scenario, though some extreme version of the SD scenario, involving high mass loss from a red giant, might work as well. There are some indications that SNe~Ia-CSM may all be of the relatively bright SN~1991T type.

The general impression that one gets from this discussion of the minority subclasses is that it is conceivable that many (perhaps most) of these objects are the results of the SD (or possibly the CD) scenario. If confirmed, this would mark a fairly significant step forward in elucidating the nature of the progenitors of SNe~Ia, even though it would still leave the majority, the more ``normal'' SNe~Ia, to be explained.

We shall now present our tentative conclusions regarding the normal SNe~Ia.

\section{Conclusions\label{sec5}}
As the current review clearly demonstrates, \textit{all} the existing progenitor scenarios encounter difficulties, if viewed as comprehensive models for all Type~Ia supernovae. In fact, partisans of particular models have been very successful in criticizing models other than their own. On one hand, this could be regarded as healthy scientific skepticism. On the other, it raises the question of whether it is at all reasonable to expect a single scenario to explain all SNe~Ia, or to even be the dominant scenario.

An examination of the observational data suggests that the bulk of the ``normal'' SNe~Ia are similar in nature (even if forming something akin to a continuum in properties). At the same time, many of the peculiar events appear to be peculiar in their own way. It therefore makes sense to examine whether a particular progenitor scenario dominates in the production of most (if not all) of the normal SNe~Ia.

On the face of it, the single degenerate scenario appears to provide for a very natural way to grow a WD to the Chandrasekhar limit and to result in an explosion through the delayed detonation (deflagration to detonation) mechanism. Past objections that claimed that the WD cannot grow in mass through accretion appear to have been overcome through the fact that heating of the WD results in weaker flashes with no significant mass loss. The SD scenario (with explosion at the Chandrasekhar limit) can account for a few of the important characteristics of SNe~Ia: (i)~The presence of stable nuclear-statistical-equilibrium material; (ii)~The width-luminosity relation in the light curve(it is reproduced at least in 1D models); (iii)~The near spherical symmetry and lack of polarization.

At the same time, the explanations suggested for the apparent absence of a companion in several cases appear rather contrived (even though not entirely impossible). While the SD scenario does not provide for a very natural explanation for the Delay Time Distribution, it is not obvious that it cannot explain it, at least within the uncertainties that are associated with both evolutionary calculations and the observations.

By comparison, it appears to us that the double-degenerate scenario (in its different guises) may leave more questions unanswered, if taken as the dominant channel for normal SNe~Ia. For example, it is not absolutely clear that it can reproduce the width-luminosity relation, it introduces non-spherical symmetry and it may be expected to show polarization. At the same time, it can naturally produce the DTD and it is consistent with the absence of detectable companions.

An interesting question that may be asked about each one of the SD, CD and the DD scenarios is the following: \textit{If for some reason any one of these scenarios does not produce a Type~Ia supernova, what does it produce?}  The impression is that a smooth merger of two WDs (where one is dissolved to form a disk around the primary) can lead to an accretion-induced collapse that forms a neutron star. This would not be the expected result in the case of a violent merger or a collision (e.g., induced by a third body). If the total mass of the two WDs is smaller than the Chandrasekhar mass, a merger could perhaps lead to subluminous events such as SN~1991bg.

Single degenerates could perhaps lead occasionally to pure deflagrations and events such as SN~2002cx (or the so-called SNe~Iax).

Overall, given the fact that serious difficulties are still associated with all scenarios, \textit{it is difficult to escape the conclusion that even just the normal SNe~Ia may be produced by a mixture of SD and DD (of various ``flavors'') perhaps even with traces of CD and other channels.} This situation is reminiscent of the observations of Gamma-ray bursts, before the realization that there are relatively long bursts (longer than about two seconds) and short bursts (shorter than two seconds). We now know that the long bursts result from the collapse of the cores of massive stars, while the short bursts are produced by neutron star--neutron star (or neutron star--black hole) collisions (the most recent confirmation coming from observations of gravitational waves and an electromagnetic follow-up; Abbott et~al. 2017b; LIGO and Virgo collaborations \textit{Fermi} and INTEGRAL 2017). Since processes involving thermonuclear explosions tend to be astrophysically messier (and also the ratio of the distance scales between the SD and DD configurations is much smaller than the ratio between the scale of a giant star and of two merging neutron stars), it may be more difficult to disentangle the different scenarios in the SNe~Ia case. 

To date, the delayed detonations of SDs and associated phenomena have been studied at greater depth than other scenarios. The head-on collisions of two white dwarfs have some promising features, and upcoming results from the \textit{Gaia} space observatory will clarify whether the statistics of stellar triples are such that collisions induced by the interaction with a third body can provide for a significant contribution to the SNe~Ia rate. Overall, SNe~Ia may represent an embarrassment of riches, with a number of classes of progenitors being responsible for these spectacular cosmic explosions.

\section*{Acknowledgements}
	
	We are grateful to David Branch, Avishay Gal-Yam, Yael Hillman, Boaz Katz, Doron Kushnir, Dan Maoz, Adam Riess,  Ofer Yaron, and in particular to Noam Soker, for very fruitful discussions. We also thank an anonymous referee for his/her careful reading of the manuscript and for detailed comments that helped in improving the text.




	
\end{document}